%% file: main.tex
\documentclass{lmcs}
\pdfoutput=1

\usepackage{lastpage}
\lmcsdoi{21}{2}{13}
\lmcsheading{}{\pageref{LastPage}}{}{}%
{Dec.~01,~2023}{May~16,~2025}{}


\usepackage[T1]{fontenc}
\usepackage[utf8]{inputenc}
\usepackage{cancel}
\usepackage{yfonts}
\usepackage{amssymb}
\usepackage{amsfonts}
\usepackage{amsthm}
\usepackage{mathtools}
\usepackage{calc}
\usepackage{hyperref}
\usepackage{graphicx}
\usepackage{multicol}
\usepackage{dsfont}
\usepackage{amsmath}
\usepackage{stmaryrd}
\usepackage{verbatim}
\usepackage{blkarray}
\usepackage{proof}
\usepackage{scalerel}
\usepackage{comment}
\usepackage[normalem]{ulem}

\usepackage{subcaption}

\usepackage{tikz}
\usetikzlibrary{cd}

\usetikzlibrary{calc,shapes,arrows,positioning,automata}

\usepackage{etoolbox}

\newtoggle{extern}
\togglefalse{extern}

\DeclareUnicodeCharacter{27E9}{$\rangle$}

\input{macros.tex}

\allowdisplaybreaks

\makeatletter
\DeclareFontEncoding{LS2}{}{\noaccents@}
\makeatother
\DeclareFontSubstitution{LS2}{stix}{m}{n}
\DeclareSymbolFont{largesymbolstix}{LS2}{stixex}{m} {n}
\DeclareMathDelimiter{\lParen}{\mathopen}{largesymbolstix}{"DE}{largesymbolstix}{"02}
\DeclareMathDelimiter{\rParen}{\mathclose}{largesymbolstix}{"DF}{largesymbolstix}{"03}

\begin{document}

\title[The Many-Worlds Calculus]{The Many-Worlds Calculus}

\author[K.~Chardonnet]{Kostia Chardonnet\lmcsorcid{0009-0000-0671-6390}
}[a,d]

\author[M.~De Visme]{Marc de Visme\lmcsorcid{0009-0004-7227-7540}
}[b]

\author[B.~Valiron]{Benoît Valiron\lmcsorcid{0000-0002-1008-5605}
}[c]

\author[R.~Vilmart]{Renaud Vilmart\lmcsorcid{0000-0002-8828-4671}
}[b]

\address{Department of Computer Science and Engineering, University of Bologna, Italy.}
\email{kostia.chardonnet@pm.me}

\address{Université Paris-Saclay, CNRS, ENS Paris-Saclay, Inria, Laboratoire Méthodes Formelles, 91190, Gif-sur-Yvette, France.}
\email{\{mdevisme, rvilmart\}@lmf.cnrs.fr}

\address{Université Paris-Saclay, CNRS, CentraleSupélec, ENS Paris-Saclay, Inria, Laboratoire Méthodes Formelles, 91190, Gif-sur-Yvette, France.}
\email{benoit.valiron@universite-paris-saclay.fr}

\address{Université de Lorraine, CNRS, Inria, LORIA, F-54000 Nancy, France}

\begin{abstract}
  In this paper, we explore the interaction between two monoidal
  structures: a multiplicative one, for the encoding of pairing, and
  an additive one, for the encoding of choice. We propose a colored PROP to
  model computation in this framework, where the choice is
  parameterized by an algebraic side effect: the model can support
  regular tests, probabilistic and non-deterministic branching, as
  well as quantum branching, i.e. superposition.

  The graphical language comes equipped with a denotational semantics
  based on linear maps, and an equational theory. We prove the
  language to be universal, and the equational theory to be complete
  with respect to this semantics.
\end{abstract}

\maketitle

\section{Introduction}
\label{sec:intro}

The basic execution flow of a computation is arguably based on three
notions: sequences, tuples and branches. Sequences form the building
block of compositionality, tuples are what makes it possible to
consider multiple pieces of information together, while branches allow
the behavior to change depending on the inputs or on the state of the
system.

Ranging from (sometimes informal) flow-chart
languages~\cite{Bartha1977finite} to sophisticated structures such as
interaction or proof nets~\cite{Lafont1989interaction,proofnets},
graphical languages are commonly used to represent the possible
control flow of a computation. On a formal level, a graphical language
is a PROP~\cite{Lack2004composing}, that is, a symmetric, strict
monoidal structure $(\mathcal{C},{\parallel},\varnothing)$ whose
objects are of the form $W\parallel\cdots\parallel W$, with
$\varnothing$ the unit for $\parallel$. The object $W$ is a ``wire'',
and any object stands for a bunch of wires. The monoidal structure
$\parallel$ formalizes how the bunching of wires behaves. We note that
the usual symbol for a monoidal structure is $\otimes$, but for
clarity we will avoid using it in situations where it is not the usual
tensor product.

\medskip
\noindent
\textbf{Two Canonical Monoidal Structures.}~
The monoidal structure of a PROP is very versatile. On one hand, it
can be considered in a \emph{multiplicative} way, with $A\parallel B$
seen as the pairing of an element of type $A$ and an element of type
$B$. This approach is one followed in the design of MLL proof-nets for
instance~\cite{goisync,proofnets}. On the other hand, one can consider
the monoidal structure in an \emph{additive} way, with $\parallel$ for
instance being a co- or a bi-product. Standard examples are the
category \cat{FinRel} of finite sets and relations, forming an
additive PROP with $\parallel$ being the disjoint union, or the
category of finite dimensional vector spaces (or
semimodules\footnote{We consider \emph{free} finite dimensional
semimodules, that is semimodules that have a finite basis, which allow
a matricial representation of morphisms, and not just a finite
generating set.}) and linear maps, with $\parallel$ being the
cartesian product.
From a computational perspective, an additive monoidal structure can
be regarded as the possibility to \emph{choose} a computational path
upon the state of the input. Depending on the underlying system, this
choice can be regarded as deterministic (if based on Set),
non-deterministic (if based on Rel), probabilistic (if based on a
suitable semimodule), \textit{etc}.

To be able to handle both pairing and branching in a PROP, we cannot
uniquely identify $\parallel$ as being multiplicative additive. We
instead need to \emph{extend} the PROP with two additional
structures, one for pairing\footnote{Once again we avoid the usual symbol $\otimes$ and reserve it for situation in which it is the usual tensor product.} (${\tensor}$) and one for branching
(${\oplus}$).

\begin{figure}[t]
  \begin{subfigure}[b]{0.43\textwidth}
    \includegraphics[width=\textwidth,page=1]{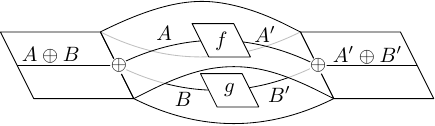}
    \caption{Split over Coproduct}
    \label{fig:intro-ex-1}
  \end{subfigure}
  \hfill
  \begin{subfigure}[b]{0.43\textwidth}
    \includegraphics[width=\textwidth]{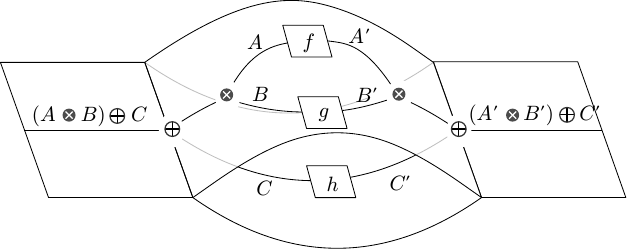}
    \caption{Splits over Coproduct and Tensor}
    \label{fig:intro-ex-2}
  \end{subfigure}
  \caption{Examples of Branchings}
\end{figure}

In this paper, we focus on a framework where these two monoidal
structures are available. Graphical languages for such a setting
usually rely on a notion of \emph{sheet}, or \emph{worlds}, to handle
general branching~\cite{duncan2009generalized,mellies2014local}.
Figure~\ref{fig:intro-ex-1} shows for instance how to represent the
construction of the morphism $f\oplus g : A\oplus B\to A'\oplus B'$
out of $f:A\to A'$ and $g:B\to B'$. The symbol ``$\oplus$'' stands for
the ``split'' of worlds. Such a graphical language therefore comes
with two distinct ``splits'': one for the monoidal structure ---living
inside one specific world---, and one for the coproduct ---splitting
worlds---. They can be intertwined, as shown in
Figure~\ref{fig:intro-ex-2}. Another approach followed by
\cite{comfort2020sheet} externalizes the two products (tensor product
and coproduct) into the structure of the diagrams themselves, at the
price of a less intuitive tensor product and a form of synchronization
constraint.

However, in the state of the art this ``splitting-world''
understanding has only been carried for deterministic or probabilistic
branching~\cite{goisync,duncan2009generalized,staton2015algebraic}. These
existing approaches do not support more exotic branchings, such as
\emph{quantum superposition}.

\medskip
\noindent
\textbf{Quantum Computation.}~
Conventional wisdom has it that quantum computation is about
\emph{quantum data in superposition}. In the standard model, the
memory holding quantum data is encapsulated inside a coprocessor
accessed through a simple interface: The coprocessor holds
individually addressable registers consisting of \emph{quantum} bits, on
which one can apply a fixed set of operations ---\emph{gates}, e.g.~ the \emph{CNot} gate \tikzfig{CNot} that can be applied on pairs of qubits---
specified by the interface. If some of these gates can generate
superposition of data, this is kept inside the coprocessor and opaque
to the programmer. A typical interaction with the coprocessor is a
purely classical sequence of elementary operations of the form ``Apply
gate X to register $n$; apply gate Y to register $m$;
\textit{etc}''. Such a sequence of instructions is usually represented
as a \emph{quantum circuit}. In this model, a quantum program is then
a conventional program building a quantum circuit and sending it as a
batch-job to the coprocessor.

From a semantical perspective, the \emph{state} of a quantum memory
consisting of $n$ quantum bits is a vector in a $2^n$-dimensional
Hilbert space. A (pure) quantum circuit is a linear, sequential description
of elementary operations describing a \emph{linear, unitary map} on
the state space.

Coming all the way from Feyman's diagrams~\cite{feynman1965quantum},
graphical languages are commonly used for representing quantum
processes. Whether directly based on quantum
circuits~\cite{green2013quipper,goisync,paykin2017qwire,chareton-qbricks}
or stemming from categorical analysis such as the
ZX-calculus~\cite{coecke2017picturing}, these formal languages are
still tied to the quantum coprocessor model in the sense that the only
monoidal structure that can be applied to quantum information is the
(multiplicative) Kronecker product. The only possible branching is
based on the (probabilistic) measurement.

\medskip
\noindent
\textbf{Quantum Control Flow.}~
However, quantum computation is not always reducible to the quantum coprocessor model, and superposition can be generalized to computation, yielding
\emph{non-causal execution paths}. Indeed, in general, the quantum
computational paradigm features two seemingly distinct notions of
control structure. On the one hand, a quantum program follows
\emph{classical} control: it is hosted on the conventional computer
governing the coprocessor, and can therefore only enjoy loops, tests
and other regular causally ordered sequences of operations. On the
other hand, the lab bench turns out to be more flexible than the rigid
coprocessor model, permitting more elaborate \emph{purely quantum}
computational constructs than what quantum circuits or ZX-calculus
allow.

The archetypal example of a quantum computational behavior
hardly attainable within quantum circuits or ZX-calculus is the
\emph{Quantum Switch} \cite{chiribella2013quantum}. Consider two quantum bits $x$ and $y$ and two
unitary operations $U$ and $V$ acting on $y$. The problem consists in
generating the operation that performs $UV$ on $y$ if $x$ is in state
$\ket{0}$ and $VU$ if it is in state $\ket{1}$, all while making a single
use of $U$ and a single use of $V$. As $x$ can be in
superposition, in general the operation is then sending
$(\alpha\ket{0}+\beta\ket{1})\otimes \ket{y}$ to
$\alpha\ket{0}\otimes (UV\ket{y}) + \beta\ket{1}\otimes(VU\ket{y})$.
While such an operation is physically
  realizable \cite{procopio2015experimental,taddei2021computational}, it requires
  to move away from the standard quantum circuit-model \cite{chiribella2013quantum}. It
is a \emph{purely quantum test}: not only can we have values in
superpositions (here, $x$) but also \emph{execution orders}. This is allowed by physical components that have no counterpart in the quantum coprocessor model,
such as the \emph{polarizing beam splitter}, represented by $\tikzfig{pbs-gate}$.

Computational models supporting superpositions of execution orders
form an active subject of research in the literature. One strand of research consists
in proposing a suitable extension of quantum
circuits~\cite{chiribella2008transforming,portmann2017causal,vanrietvelde2021routed,wechs2021quantum}.
These approaches typically aim at discussing the notion of quantum
channel from a quantum information theoretical standpoint.
Another strand is to focus on the additive aspect of
quantum computation as in \cite{pbs} which focusses on the polarizing beam splitter. However, here, similarly to
the coprocessor model but with reversed roles, it is the multiplicative
part that is hidden inside the primitives.

\medskip
\noindent
\textbf{Limitation of Current Approaches and Objective of the Paper.}~
Although there is a finer and finer understanding of superposition of
causal orders in the literature, none of the existing PROPs can
support the quantum switch on complex data built from tensors
and coproducts at a fine-grain level, which becomes necessary when trying to integrate indefinite causal orders to usual tensor-based models of quantum computation. Our aim in this paper is to give an adequate framework
for understanding non-ambiguously diagrams such as:
\[\tikzfig{example-mixing-plus-tensor}\]
By doing so, we shall be able to express operators composed of
polarising beam splitters, but where (i) the controlled system is explicit,
and there are no black boxes as primitives, and (ii) the controlled system
can become the control system later on, and vice-versa; harnessing the
power of both indefinite causal order-enabling primitives and
superposition-inducing quantum gates. Having fine-grained primitives
both quantum data and quantum control flow moreover shall permit better
implementation capabilities and automated reasoning on programs making use
of them.

We claim in this paper that the same intuition
underlying probabilistic branching can be followed for quantum (and
more general) branching. In the conventional case, $1\oplus 1$ is a
regular boolean: either ``left'' (standing e.g. for True) \emph{or}
``right'' (standing e.g. for False).
In quantum computation, the sum-type $1\oplus 1$ can however be
understood as a \emph{sum of vector spaces}, giving an alternative
interpretation to $1\oplus 1$: it can be regarded as the type of a
quantum bit, superposition of True and False. One should note that
this appealing standpoint should be taken cautiously: (Pure) quantum
information imposes strong constraints on the structure of the data
in superposition: orthogonality and unit-norm have to be
preserved~\cite{grattage05functional,sabry2018symmetric}.

\begin{figure}[t]
  \centering
  \includegraphics[width=.48\textwidth]{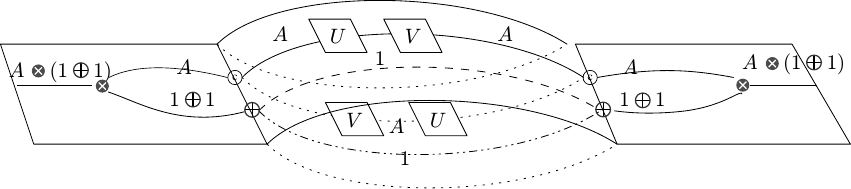}
  \caption{Quantum Switch with Worlds}
  \label{fig:intro-q-switch-sheet}
\end{figure}

The Quantum Switch can then be naturally understood in this framework.
Consider for instance Figure~\ref{fig:intro-q-switch-sheet}, read from
left to right: as input, a pair of an element of type $A$ and a
quantum bit. Based on the value of the qubit (True or False), the wire
$A$ goes in the upper or the lower sheet, and is fed with $U$ then $V$
or $V$ then $U$. Then everything is merged back together.
However, while in Figure~\ref{fig:intro-q-switch-sheet} two copies of
$U$ and $V$ are required, we will see that in our system only one copy
of each is needed.

\smallskip
\noindent
\textbf{Contributions.}~
In this paper, we introduce a new graphical language for quantum
computation, based on compact category with biproduct
\cite{Heunen2019categories}. This language allows us to express any
process with both pairing and a general notion of algebraic branching,
encompassing deterministic, non-deterministic, probabilistic and
quantum branching. We develop a denotational semantic and an
equational theory, and prove its soundness and completeness with
respect to the semantics. As a case-study, we show how the Quantum
Switch can naturally be encoded in the language as well as the quantum
programming language for quantum control presented
in~\cite{sabry2018symmetric}. Additionally, while in
Figure~\ref{fig:intro-q-switch-sheet}, we represent the quantum switch
with two syntactic copies of $U$ and $V$, this is merely a pedagogical
choice. As shown in Figure~\ref{fig:rewrite_QS}, we are able to
represent the quantum switch with only one syntactic instance of $U$
and $V$, allowing us to write an higher-order version of this Quantum
Switch in Figure~\ref{fig:QS_higher}.

\section{The Many-Worlds Calculus}
\label{sec:many-worlds}
Our calculus is parameterized by a commutative semiring
$(R, +, 0, \times, 1)$. It can be instantiated by the complex numbers
$(\mathbb{C}, +, 0, \times, 1)$ to represent pure quantum computations,
the non-negative real numbers $(\mathbb{R}_{\geq 0}, +, 0,\times, 1)$ for
probabilistic computations, or the booleans
$(\{\text{False},\text{True}\},$ $ \text{OR}, \text{False}, \text{AND}, \text{True})$
for non-deterministic computations.

The goal is to define a graphical language in which each wire can be
enabled or disabled depending on the world in which the computation
takes place.To do so, we introduce three categories (actually
colored PROPs), each of which being a refinement on the previous one
and addressing its main caveat. These three categories are recalled in
the following table, for reference, but are developed in details in
the rest of the section.
\begin{table}[!htb]
\begin{tabular}{|p{4.2cm}|p{5.25cm}|p{4.5cm}|}
\hline
\multicolumn{1}{|c|}{$\catD$} &  \multicolumn{1}{|c|}{$\ManyWorlds{W}$} &  \multicolumn{1}{|c|}{$\ManyWorlds{\forall}$}\\
&\multicolumn{1}{|c|}{for W a set of worlds}&
\\
\hline
\emph{Colors:}\newline$\begin{array}{lrl@{~}l@{~}l} A, B &::= & \zero & \mid & A\oplus B \\& \mid & \one & \mid& A\tensor B \end{array}$ &
\emph{Colors:}\newline$\begin{array}{@{}l@{}r@{}} (A:w) \text{ with } & A \in \operatorname{Colors}(\catD),\\& w \subseteq W\end{array}$&
\emph{Colors:}\newline same as \catD
\\[0.5em]
\emph{Objects:}\newline$\begin{array}{lrl} \mathcal{A} &::= & \emptysquare \\ &\mid & A_1 \parallel \dots \parallel A_n\end{array}$&
\emph{Objects:}\newline$\begin{array}{@{}l@{}l@{}}(\mathcal A, \ell_{\mathcal A}) & \text{ with } \mathcal A\in \operatorname{Objects}(\catD),\\& \ell_{\mathcal A}:\{A_i\in\mathcal A\}\to\mathcal P(W)\end{array}$&
\emph{Objects:}\newline same as \catD
\\[0.5em]
\emph{Morphisms:}\newline sequential and parallel \rule{0ex}{0ex} composition of elements Figure~\ref{fig:generator}.&
\emph{Morphisms:}\newline$\begin{array}{@{}l@{}l@{}} (f, \ell_{f}) & \text{ with } f\in \catD,\\&\ell_{f}:\operatorname{Wires}(f)\to\mathcal P(W)\end{array}$&
\emph{Morphisms:}\newline$\withWorld{f}{W}\in\operatorname{Morphisms}(\ManyWorlds{W})$
\\[0.5em]
\emph{Compositions:}\newline usual sequential and parallel composition of colored PROP.&
\emph{Compositions:}\newline usual sequential and parallel composition of colored PROP.&
\emph{Compositions:}\newline$\begin{array}{@{}l@{}l@{}}\withWorld{f}{W} \parallel \withWorld{g}{V} & ~:=\\&\withWorld{(f^{- \times V} \parallel g^{W \times -})}{W \times V} \end{array}$
\newline$\withWorld{g}{V} \circ	\withWorld{f}{W} :=$\newline\quad $\withWorld{(g^{(W \times -)\cap Z} \circ f^{(- \times V) \cap
		Z})}{Z}$\\
\hline
\end{tabular}
\label{tab:overview-of-props}
\caption{Overview of the introduced colored PROPs.}
\end{table}

\subsection{The Unlabeled Language}

We define our graphical language within the paradigm of colored PROP
\cite{Carette21,HackneyR15}, meaning that a morphism is a diagram composed
of nodes, or \emph{generators}, linked to each other through
\emph{colored wires}, that are allowed to cross each
others. Additionally, we assume that our colored PROP is compact
closed and auto-dual, i.e.~we allow to bend wires to obtain a
Cup (\scalebox{0.8}{$\tikzfig{lang/cup}$}) or a Cap (\scalebox{0.8}{$\tikzfig{lang/cap}$}).

The generators of our language are described in Figure~\ref{fig:generator}
and are respectively the Identity, the Swap, the Cup, the Cap, the
Plus, the Tensor, the Unit, the $n$-ary Contraction for $n \geq 0$,
and the Scalar for $s$ ranging over the commutative semiring $R$.
Mirrored versions of those generators are defined as syntactic sugar
through the compact closure, as shown for the mirrored Plus on the
right-hand-side of Figure~\ref{fig:generator}.  Diagrams are read
top-to-bottom: the top-most wires are the \textit{input} wires and the
bottom-most wires are the \textit{output} wires. The labels $A$, $B$,
etc correspond to the colors of our PROP. There is one color for every
type generated by the syntax\footnote{Those types are purely
syntactic. We do not assume any associativity/distributivity/etc at
the syntactic level. However, we have morphisms representing those
properties (e.g. a morphism from $(A \oplus B) \oplus C$ to $A \oplus
(B \oplus C)$), and up to the equational theory $\equiv$ those
morphisms will be isomorphisms.} $A, B ::= \one \mid \zero \mid A\oplus B \mid
A\tensor B$. As such, the Unit starts a wire of type $\one$, the Plus
combines two wires of type $A$ and $B$ into a wire of type $A \oplus
B$, and similarly the Tensor combines  two wires of type $A$ and $B$
into a wire of type $A \tensor B$.

In a colored PROP, an object $\mathcal{A}$ is simply a list of colors
$\mathcal{A} = A_1 \parallel \dots \parallel A_n$ (or $\mathcal{A} =
\emptysquare$ for the empty collection). For example, the Plus is a
morphism from $A \parallel B$ to $A \oplus B$. The choice of the
notation $\parallel$ for wires in parallel is uncommon, we use it to put
an emphasis on the fact that contrary to languages like the
ZX-calculus, wires that are in parallel are not necessarily ``in tensor
with one another''. In fact, $A \parallel B$ can be understood
semantically as ``either $A \tensor B$ or $A \oplus B$''.

Diagrams are obtained from generators given in Figure~\ref{fig:generator}
and by composing them in parallel (written $\parallel$), or
sequentially (written $\circ$). Sequential composition requires the
type (and number) of wires to match. Notice that there is no
generator for the type $\zero$, as it is a special case of the
contraction with $n = 0$ and on type $\zero$, i.e. $\tikzfig{lang/zero-inv}$. We write \catD~for the category
of diagrams we defined as such, and $\catD(\mathcal{A},\mathcal{B})$
for the set of diagrams that are morphisms from $\mathcal{A}$ to
$\mathcal{B}$.

    \[ D_2\circ D_1:=\tikzfig{compo-seq}\qquad D_1\parallel
D_2:=\tikzfig{compo-ten} \]

\begin{figure}[t]
	$\tikzfig{lang/id}  \quad  \tikzfig{lang/swap} 		 \quad
	\tikzfig{lang/cup}  \quad  \tikzfig{lang/cap} 	     \quad
	\tikzfig{lang/plus} \quad \tikzfig{lang/tensor-PN}   \quad
	\tikzfig{lang/unit} \quad \tikzfig{lang/contraction} \quad
	\quad \tikzfig{lang/scal}$					 		 \\[0.5em]
	$\tikzfig{lang/plus-inv} := \tikzfig{lang/plus-cc} $
	\caption{Generators of our Unlabeled Language ($n \geq 0$, $s \in R$)}
	\label{fig:generator}
\end{figure}

\begin{exa}
	We consider $R = \mathbb{C}$. While \catD~lacks the worlds
	labeling\footnote{We call worlds labeling the attribution of
	\emph{multiple} worlds to a single wire, as defined in
	Section~\ref{sec:worlds}.}, we can already illustrate our language by
	encoding some basic quantum primitive in it and show how they operate.
	In Figure~\ref{fig:hadamard} we show the encoding of a quantum bit $\alpha
	\ket{0} + \beta \ket{1}$ and the Hadamard unitary. In particular, the
	Plus allows to ``build'' a new quantum bit from two scalars in parallel
	or to ``open'' a quantum bit to recover its corresponding scalars, the
	left branch corresponding to $\ket{0}$ and the right branch to
	$\ket{1}$. The meaning of the Contraction is better seen when applying
	Hadamard to a quantum bit as we show in Figure~\ref{fig:rewrite}: it allows
	us to duplicate and sum scalars. The rewriting sequence of
	Figure~\ref{fig:rewrite} is made using the equational theory defined in
	Section~\ref{sec:equation}, however to correctly define our equational
	theory, the worlds labeling are required. So while this specific
	worlds-free rewriting sequence is sound, many other similar
	worlds-free rewriting sequences are unsound.
\end{exa}

\begin{figure}[!htb]
\begin{align*}
\alpha \ket{0} + \beta \ket{1} &\rightsquigarrow~ \tikzfig{oneqbit} \in \catD(\emptysquare,\one \oplus \one)\\
\begin{pmatrix}
	\frac{1}{\sqrt{2}} & \frac{1}{\sqrt{2}} \\\frac{1}{\sqrt{2}} & \frac{-1}{\sqrt{2}}
	\end{pmatrix}&\rightsquigarrow~ \tikzfig{hadamard} \in \catD(\one \oplus \one,\one \oplus \one)
\end{align*}
	\caption{A Quantum Bit and the Hadamard Unitary}
	\label{fig:hadamard}
\end{figure}

\begin{figure}[!htb]
	$\tikzfig{h-oneqbit-1} \quad \to \quad
	\tikzfig{h-oneqbit-2}\to\quad\tikzfig{h-oneqbit-3}\quad\to\quad\tikzfig{h-oneqbit-4}$
	\caption{Applying the Hadamard Unitary to a Quantum Bit}
	\label{fig:rewrite}
\end{figure}

\begin{rem}
	\label{rem:cupcap}
	Instead of having the Cup and the Cap as generators and
        defining the mirrored version of each generator through them,
        one could proceed the other way around by defining the Cap as
        follows, and the Cup in a mirrored
        way:\\ $\phantom{.}\hfill\tikzfig{cupcap/tens} \hfill
        \tikzfig{cupcap/plus} \hfill
        \tikzfig{cupcap/one}\hfill\phantom{.}$
\end{rem}

\subsection{Sheets and 3 Dimensional Diagrams}

We said that $A \parallel B$ corresponds to ``either $A \oplus B$ or $A
\tensor B$''. There are circumstances, in particular when trying to
define an equational theory, where it is useful to know which one of
the two it is. Following \cite{comfort2020sheet}, we could draw our
diagrams on sheets, leading to a 3D figures like
Figures~\ref{fig:intro-ex-1}, \ref{fig:intro-ex-2} and~\ref{fig:intro-q-switch-sheet}. In
those figures, wires that are in parallel sheets are $\oplus$-related
while wires that are in the same sheets are $\tensor$-related.
Additionally, the nodes $\oplus$ and $\contraction$ allow for parallel
sheets to be merged together, while every other generator has no
effect on the sheet structure.

\begin{figure}
	\[\tikzfig{not-sheets}\]
	\caption{The Negation from $\one \oplus \one$ to $\one \oplus \one$.}
	\label{fig:negation_cross}
\end{figure}

However, drawing in 3D can quickly get messy, especially when one needs
to make sheets cross each others (see Figure~\ref{fig:negation_cross}). A
compromise is to instead annotate each wire by the name of the sheet
it is on, so for example $\tikzfig{explicit/plus-w}$ would mean that
we have an $\oplus$ node merging the sheet $w$ and the sheet $v$ into
the sheet $w \sqcup v$. Going even further, we can rewrite the notion
of sheets into a notion of worlds:
\begin{itemize}
	\item A sheet corresponds to a set of worlds. We annotate every
	wire of the sheet by its world set. Wires that are
	$\tensor$-related have the same annotation.
	\item Parallel sheets are disjoint sets of worlds. Wires that are
	$\oplus$-related have annotations that are disjoints sets of
	worlds. Merging two sheets $w$ and $v$ correspond to a disjoint
	union $w \sqcup v$.
	\item Wires that are world sets that are neither equal or disjoint
	are neither $\tensor$-related or $\oplus$-related, but are instead
	of mixture of both. One notable case where this situation happen
	is when decomposing a wire of type $(A \oplus B) \tensor (C \oplus
	D)$ into four wires of types $A,B,C$ and $D$, as in
	Figure~\ref{fig:neither_oplus_or_tensor}. The possibility to have world
	sets in this situation also allows us to have a very lax compact
	structure, where wires can be bent at will, e.g.~in
	\tikzfig{CEX-fully-typed-morphism}
\end{itemize}

\begin{figure}
	\[\tikzfig{explicit/NeitherPlusTensor}\]
	\caption{The Wires of Type $B$ and of Type $C$ are Neither $\tensor$-Related or $\oplus$-Related.}
	\label{fig:neither_oplus_or_tensor}
\end{figure}

\subsection{Adding Worlds Labeling}
\label{sec:worlds}
We now label wires of our diagram with sets of worlds $w \subseteq W$
from a given world set $W$. For each world $a \in W$, wires labeled
by a set containing $a$ are said to be ``enabled in $a$'', and the
others are said ``disabled in $a$''. This allows us to correlate the
enabling of wires. Before making this formal, we illustrate this
notion through the following example:

\begin{figure}[!htb]
	\[\tikzfig{explicit/CNOT-Worlds}\]
	\caption{Controlled Not with World Set $\{ \aa,\bb, \cblue, \sstar\}$}
	\label{fig:CNOT}
\end{figure}

\begin{exa}\label{ex:CNOT} The ``Controlled Not'' on quantum bits
can be represented by the Figure~\ref{fig:CNOT}. The figure is split into
two parts: the control part on the left-hand-side, and the
computational part on the right-hand-side.  The idea is that the
control part, that uses $\oplus$, will behave as an
\textit{if-then-else} and will bind the world $\aa$ to the case where
the control quantum bit is $\ket{0}$, and the worlds $\bb$ and
$\cblue$ to the case where the control quantum bit is  $\ket{1}$.
Lastly, the world $\sstar$ appears nowhere in the labels, and
corresponds to ``we do not evaluate this circuit at all''\footnote{While
not strictly necessary, it is often practical to have a world absent
from every wire. We discuss its usefulness in
Remark~\ref{rem:star_worlds}.}. On the computational part, we apply the
identity within the world $\aa$, we negate $\ket{0}$ into $\ket{1}$
within $\bb$, and we negate $\ket{1}$ into $\ket{0}$ within $\cblue$.
The domain and codomain of this labeled diagram are $\left(\one \oplus
\one: \{ \aa,\bb, \cblue\}\right) \parallel \left(\one \oplus \one :
\{ \aa,\bb, \cblue\}\right)$, which we will write
$(\mathcal{A},\ell_{\mathcal{A}})$ with an object $\mathcal{A} = (\one
\oplus \one)\parallel (\one \oplus \one)$ and a labeling function
$\ell_{\mathcal{A}} : 1 \mapsto  \{ \aa,\bb, \cblue\} \quad 2 \mapsto
\{ \aa,\bb, \cblue\} $. Similarly, the ``Controlled Not'' above can be
seen as a diagram $\D_{\text{CNOT}}$ of
$\catD(\mathcal{A},\mathcal{A})$ together with a labeling function
$\ell_{\mathcal{A}}$ which labels every wire with a set of worlds.
\end{exa}

We now give a formal definition of the concept of worlds:
\begin{defi}\rm
  Given a set of worlds $W$, we define the auto-dual compact closed
  colored PROP $(\ManyWorlds{W},\parallel,\emptysquare)$ of
  Many-Worlds calculus over $W$ as follows:

  \smallskip
  \noindent
  \emph{Its colors} are the pairs $(A:w)$ of colors $A$ of $\catD$ and
  subsets $w \subseteq W$. We write $(\mathcal{A},\ell_{\mathcal{A}})$ for
  the objects, where $\mathcal{A}$ is an object of $\catD$ and $\ell_{\mathcal{A}}$
  is a labeling function from the colors\footnote{If the same color appears multiple times in the object, each instance might have a different label.} composing $\mathcal{A}$ to the subsets  of $W$.

  \smallskip
  \noindent
  \emph{Its morphisms} $f \in
  \ManyWorlds{W}((\mathcal{A},\ell_{\mathcal{A}}),(\mathcal{B},\ell_{\mathcal{B}}))$
  are pairs $(\D_f,\ell_f)$ of a morphism $\D_f \in
  \catD(\mathcal{A},{\mathcal{B}})$ and a labeling function $\ell_f$
  from the wires of $\D_f$ to the subsets of $W$, satisfying the
  following constraints: The label on an input or output wire of color
  $(A:w)$ must be equal to $w$, and\\
  $\tikzfig{explicit/id-w}\hfill\tikzfig{explicit/swap-w}\hfill\tikzfig{explicit/cap-w}\hfill\tikzfig{explicit/cup-w}
    \hfill\tikzfig{explicit/scalar-w} \hfill
    \tikzfig{explicit/plus-w}\hfill\tikzfig{explicit/tensor-PN-w}\hfill\tikzfig{explicit/tensor-unit-w}\hfill\tikzfig{explicit/contraction-w}$
    \\[3pt]
     for all $n\geq 0$ and where $\sqcup$
    denotes disjoint set-theoretic union. The constraints for the
    mirrored versions are similar.  The sequential composition $\circ$
    and the parallel composition $\parallel$ preserve the labels.

\end{defi}

We write $\withWorld{f}{W} : \mathcal{A} \to \mathcal{B}$ for $f = (\D_f,\ell_f)$
a morphism of
$\ManyWorlds{W}((\mathcal{A},\ell_{\mathcal{A}}),(\mathcal{B},\ell_{\mathcal{B}}))$,
where $\ell_{\mathcal{A}}$ and $\ell_{\mathcal{B}}$ can be deduced
from $\ell_f$ using the first restriction on labels. We note that when
considering $\withWorld{f}{W} : \mathcal{A} \to \mathcal{B}$ and $\withWorld{g}{W} :
\mathcal{B} \to \mathcal{C}$, there is no reason for the labels
deduced on $\mathcal{B}$ to be the same, so there is no reason for $\withWorld{(g
\circ f)}{W} : \mathcal{A} \to \mathcal{C}$ to be defined. And even when
$\withWorld{(g \circ f)}{W}$ is defined, it might not have the meaning we intend:
for example, Figure~\ref{fig:ex_double_not} shows a ``broken'' composition of
the negation with itself, as the world $\bb$ corresponds to ``the input
of the first negation is \ket{0}, its output is \ket{1}, then in an contradictory way the input of the second negation is back at \ket{0} and its
output is \ket{1}'', which does not make any computational sense. In
fact, if we use the equational theory $\equiv$ defined in
Section~\ref{sec:equation}, the semantics of this diagram collapses to zero.

In practical terms, the naive composition of $\ManyWorlds{W}$
can be used to decompose a diagram into fragments of said diagrams,
but is ill-suited to assemble independent diagrams together.

\begin{figure}
	\[\tikzfig{explicit/CNOT-CNOT} \quad \equiv \quad  \tikzfig{explicit/zero_scalar}\]
	\caption{``Broken'' Composition of two Negations}
	\label{fig:ex_double_not}
\end{figure}

\subsection{World-Agnostic Composition}

\begin{figure}[t]
	\[ \scalebox{0.9}{$\begin{matrix}
		W=\{a,\star\} && W = \{a,b,c,\star\} && W = \{a,\star\} && W = \{a,\star\} \\
		\tikzfig{explicit/id-can} && \tikzfig{explicit/swap-can} &&
		\tikzfig{explicit/cup-can} &&  \tikzfig{explicit/cap-can}
		\end{matrix}$} \] \caption{Canonical Labelings in
		$\ManyWorlds{\forall}$}
	\label{fig:can_lab}
\end{figure}

When building two diagrams $\withWorld{f}{W}$ and $\withWorld{g}{V}$, there
is no reason for the worlds sets $W$ and $V$ to be equal, but we might
still want to be able to compose them with one another, hence the need
for a new kind of composition that better handles the world sets.

\begin{defi} We define the auto-dual compact closed colored PROP
$(\ManyWorlds{\forall},\parallel,\emptysquare)$ of Many-Worlds
calculus as follows:

	\smallskip
	\noindent
	\emph{Its colors} and \emph{objects} are the same as the ones of \catD.

	\smallskip
	\noindent
	\emph{Its morphisms} from $\mathcal{A}$ to $\mathcal{B}$ are simply
	morphisms $\withWorld{f}{W} : \mathcal{A} \to \mathcal{B}$ of $\ManyWorlds{W}$ for
	any finite set $W$. See Figure~\ref{fig:can_lab} for the canonical
	labelings on the identity, swap, cup and cap. Morphisms are
	considered up to renaming of the worlds\footnote{More precisely,
	$\withWorld{(\D_f,\ell_f)}{W} = \withWorld{(\D_f,\sigma\circ \ell_f)}{\sigma(W)}$ for
	any $\sigma$ describing a bijection between $W$ and $\sigma(W)$.
	Without this quotient, we would not have $\withWorld{f}{W} \circ \id =
	\withWorld{f}{W}$.}.

	\smallskip
	\noindent
	\emph{The parallel composition} is given by $\withWorld{f}{W} \parallel \withWorld{g}{V}
	:= \withWorld{(f^{- \times V} \parallel g^{W \times -})}{W \times V}$ where
	$f^{\sigma(-)}$ has the same diagram $\D_f$ and has for labels
	$\ell_{f^{\sigma(-)}}(x) = \sigma(\ell_f(x))$.

	\smallskip
	\noindent
	\emph{The sequential composition} is given by $\withWorld{g}{V} \circ
	\withWorld{f}{W} := \withWorld{(g^{(W \times -)\cap Z} \circ f^{(- \times V) \cap
	Z})}{Z}$ where $Z \subseteq W \times V$ is the greatest subset
	such that this composition is well-defined, as illustrated by the
	following example and explained just after.
\end{defi}

\begin{figure*}[t]
	\[\tikzfig{explicit/agnostic_mono}\]
	\caption{First Example of Worlds-Agnostic Composition}
	\label{fig:ex1_agnostic_compo}
\end{figure*}

\begin{figure*}[t]\[\tikzfig{explicit/agnostic_compo}\]
	\caption{Second Example of Worlds-Agnostic Composition}
	\label{fig:ex2_agnostic_compo}
\end{figure*}
\begin{exa}[World Agnostic Composition]
	In Figure~\ref{fig:ex1_agnostic_compo}, we show the
	result of the parallel composition of $f_{\{\aa,\sstar\}} =
	\id_{A:\{\aa\}}$ and $g_{\{\bb,\sstar\}} = \id_{A:\{\bb\}}$ for
	a color $A$. The world $(\aa,\sstar)$ corresponds to the left wire
	being enabled and the right one disabled, the world $(\sstar,\bb)$
	is the opposite, the world $(\aa,\bb)$ corresponds to both wires
	being enabled and the world $(\sstar,\sstar)$ to both disabled.

	Then, in Figure~\ref{fig:ex2_agnostic_compo} we
	continue by composing with the Cup over $A:\{\cblue,\sstar\}$. To
	compute the composition, we proceed in two steps: first we handle
	the situation as if it were a parallel composition, leading to a
	diagram labeled over $W \times V \times U$, but with multiple
	contradictory labels on the wire. Then, we eliminate as many
	worlds as necessary to make those labels compatible:
	\begin{itemize}
		\item We eliminate $(\aa,\bb,\sstar)$ and
		$(\aa,\sstar,\sstar)$ which are on the \emph{left} label but
		not on the \emph{bottom} one.
		\item We eliminate $(\sstar,\bb,\cblue)$ and
		$(\sstar,\sstar,\cblue)$ which are on the \emph{bottom} label but
		not on the \emph{left} one.
		\item Looking at the \emph{right}, we eliminate
		$(\sstar,\bb,\sstar)$ and $(\aa,\sstar,\cblue)$ for similar
		reasons.
	\end{itemize}
	We eliminated six worlds, with the only remaining ones being $Z =
	\{(\aa,\bb,\cblue),(\sstar,\sstar,\sstar)\}$.
\end{exa}

\noindent The above procedure can be generalized for any $\withWorld{g}{V} \circ \withWorld{f}{W}$:
	\begin{itemize}
		\item We start by computing $g^{W \times -}$ and $f^{- \times V}$.
		\item Then we forcefully append the two diagrams as if one was composing them as morphisms of $\ManyWorlds{W \times V}$. We write $h$ for the resulting diagram, although it will rarely be a valid diagram of $\ManyWorlds{W \times V}$, as some wires might be labeled by multiple contradictory sets of worlds.
		\item For each wire $e_i$, we consider the various sets of worlds labeling it, let us name them $w_1^i,\dots,w_n^i$, and write $w^i = \bigcap_{k=1}^n w_k^i$. In order to make the different labelings of the wire $e_i$ consistent with each others, we need to remove all the worlds of $w_k^i \backslash w^i$ for all $k$.
		\item So writing $u = \bigcup_i \bigcup_{k=1}^n (w_k^i \backslash w^i)$, the restricted diagram $h \backslash u$ is a valid diagram of $\ManyWorlds{(W \times V)\backslash u}$. In fact, this diagram $h \backslash u$ is equal to the composition $(g^{W \times -} \backslash u) \circ (f^{- \times V})\backslash u$.
		\item Writing $Z = (W \times V)\backslash u$, $h \backslash u$ is also equal to $(g^{(W \times -)\cap Z} \circ f^{(- \times V) \cap Z})$. We can then take:
		\[\withWorld{g}{V} \circ
		\withWorld{f}{W} := \withWorld{(g^{(W \times -)\cap Z} \circ f^{(- \times V) \cap
				Z})}{Z}\]
		\item And by construction, $Z = (W \times V)\backslash u$ is indeed the greatest subset of $W \times V$ such that this composition is well-defined.
	\end{itemize}

\begin{figure}
	\[\tikzfig{explicit/CNOT-CNOT-wa} \]
	\caption{World-Agnostic Composition of two Negations}
	\label{fig:ex_double_not_worldagnostic}
\end{figure}

\begin{exa}[Back to the Double Negation]
	We show in Figure~\ref{fig:ex_double_not_worldagnostic} the result of
	the world-agnostic composition of the negation on the worlds set
	$\{\bb,\cblue\}$ with itself. The number of worlds grows
	significantly as all the pairs have to be considered, though the
	worlds $(\bb,\bb)$ and $(\cblue,\cblue)$ are meaningless
	computationally so the equational theory $\equiv$ allows us to
	remove them, leaving only the worlds $(\bb,\cblue)$ and
	$(\cblue,\bb)$ which correspond to the two different paths that
	the data can take through the double negation. Those two remaining
	worlds can then be merged together into a single world $\xx$ using
	that same equational theory, giving the identity morphism as a
	final result.
\end{exa}

\subsection{The Meaning of the \texorpdfstring{$\star$}{*} World}
\label{rem:star_worlds}

In most practical examples, we include a world which appears nowhere
in the worlds annotations and we denote it $\star$. This world
denotes the possibility of not evaluating the diagram at all,
for example if no input is ever received. It is particularly useful
in compositional approaches, where the diagram is in fact a fragment
of a bigger diagram, that might not be used in some circumstances. For
example Figure~\ref{fig:ex_CNOT_decomposed} represents the conditional
$\bfup{Cond}$ as well as the state $\bfup{q}^{\textup{NOT}}$ encoding
the negation, and should we compute $\bfup{Cond}_{\{\aa,\xx,\sstar\}}
\circ (\id \parallel \id \parallel
\bfup{q}^{\textup{NOT}}_{\{\bb,\cblue,\sstar\}})$ we would
obtain\footnote{After a few rewriting steps as defined in
Section~\ref{sec:equation}, including the following renaming of the worlds:
$(\aa,\sstar) \mapsto \aa$, $(\xx,\bb) \mapsto  \bb$, $(\xx,\cblue)
\mapsto \cblue$, $(\sstar,\sstar) \mapsto \sstar$.} the expected
``Controlled Not'' of Figure~\ref{fig:CNOT}, while if we removed the
$\sstar$ the semantics would collapse, the Not branch would then be
forced and we would obtain\footnote{After a few rewriting steps as
defined in Section~\ref{sec:equation}.} the result described in
Figure~\ref{fig:ex_CNOT_decomposed_recomposed}.

\begin{figure}[t]
	\[\tikzfig{explicit/CHighOrder}  \] \caption{The Conditional
	\bfup{Cond} and the State Encoding the Negation
	$\bfup{q}^{\textup{NOT}}$, with Respective World Sets
	$\{\aa,\xx,\sstar\}$ and $\{\bb,\cblue,\sstar\}$.}
	\label{fig:ex_CNOT_decomposed}
\end{figure}

\begin{figure}[t]
	\[\tikzfig{explicit/CNOT-Broken}  \]
	\caption{The ``Broken'' CNot Obtained when Composing Without the $\star$ World. Note that every world of the form $(\aa,-)$ has been eliminated during the composition, and the corresponding wires have been eliminated using the equational theory $\equiv$.}
	\label{fig:ex_CNOT_decomposed_recomposed}
\end{figure}

\begin{rem}[Quantum Conditionals]
	This behavior is already known in quantum computation:
	quantum conditionals are known to be impossible to implement
	due to some no-go theorems \cite{Araujo2014impossibleQif},
	however, some implementations actually exists
	\cite{Zhou2011possibleQif}, this apparent contradiction can
	be solved by introducing a notion of sectors
	\cite{Vanrietvelde2021sectors} to show that in the former the
	controlled operation was expected to ``always be used'' while
	in the latter it ``might or might not be used''.
\end{rem}

\begin{rem}[Supermaps and Currying]
	\label{rem:currying}
	Notice that the Many-Worlds can represent higher-order
	quantum processes, i.e. supermaps. This is done through its
	compact structure. Generaly, supermaps are
	represent with diagrams with holes (as shown below, on the
	left-hand-side) where the hole can then be filled by a diagram
	going from $A$ to $B$. By bending the wire, we can transform this
	diagram into one that takes two additional input $A$ and $B$.
	\[\tikzfig{supermaps}\] For a concrete exemple, this is what
	happens in the left-hand-side diagram of
	Figure~\ref{fig:ex_CNOT_decomposed}, where \bfup{Cond} is awaiting
	for a diagram from $\one\oplus\one \to \one\oplus\one$.

\end{rem}

\section{Representing Computation}
\label{sec:isos-language}

As a motivational example, we may see how this Many-Worlds diagrams
can be used to represent computations expressed in a language that
explicitly uses the two compositions $\tensor$ (through pairs), and
$\oplus$ (through pattern-matching), as in
\cite{sabry2018symmetric,phd-kostia}. As this translation is not the
focus of this paper, a lot of technicalities are glossed over. More
details of this translation can be found in~\cite{phd-kostia}.

\begin{figure*}[t]\scalebox{0.9}{\tikzfig{terms/exemple}}
	\centering
	\caption{Representation of the term $t$ from
		Example~\ref{ex:iso-to-MW}.}\label{fig:ex_match}
\end{figure*}
\subsection*{Syntax of the Language}

The language we present here is adapted from
\cite{sabry2018symmetric}, where we consider the values and types
together with the enrichment that is linear combinations of terms, but
without abstraction nor recursion. The syntax that is used in the
language is given as follows, with scalars $s$ ranging over the
commutative semiring $R$:
  \begin{alignat*}{100}
  \label{tab:syntax}
    &\text{(Base types)} \quad& A, B &&&::=~ && \one \alt A \oplus B
    \alt A \tensor B \\
    &\text{(Isos, first-order)} & \alpha &&&::=&&A\isot B \\[1.5ex]
    &\text{(Values)} & v &&&::=&& \langle\rangle \alt x \alt \inl{v} \alt \inr{v} \alt \pv{v}{v} \\
	&\text{(Pattern)} & p &&& ::= && x \mid \pv{p_1}{p_2} \\
    &\text{(Expressions)} & e &&&::=&& v \alt
    \letv{p}{\omega~p}{e} \alt e+e \alt \alpha e \\
    &\text{(Isos)} & \isoterm &&&::=&& \isobasique \\
    &\text{(Terms)} & t &&&::=&& \langle\rangle \alt x \alt \inl{t} \alt \inr{t} \\
    &&&&&&& \alt \pv{t_1}{t_2} \alt \letv{p}{t}{t} \\
    & & &&& && \alpha t \alt t + t \alt \isoterm~t
  \end{alignat*}

The language in particular features branching through the $\inl{}$ and
$\inr{}$ constructors, linear combinations of terms and expressions,
and crucially \emph{isos} that have type $A\leftrightarrow B$: they
turn a term of type $A$ to a term of type $B$ using pattern-matching.
The language comes with a predicate (not presented here, although it
could be given a diagrammatic meaning), used in the typing rule of
isos, to ensure exhaustivity and the non-overlapping character of the
left-hand and right-hand expressions of the clauses, allowing in particular
to define unitaries (in the complex setting). Constraints on the
linear combinations may also be used to enforce probabilistic
constraints (i.e.~that states are normalized in the quantum setting).

There are two different typing judgements, one for terms $\vdash_t$
and one specific for isos $\vdash_\omega$. In the following, we will
use the shorthands $\fc:=\inl\langle\rangle$ and
$\tc:=\inr\langle\rangle$.

\begin{exa}
\label{ex:Hadamard-in-language}
  In the case where $R = \mathbb C$, one can encode the Hadamard and
  the CNOT gate by:
  \[\begin{array}{ll}
    \opn{Hadamard}:\one\oplus\one\isot \one\oplus\one \\
    \quad = \left\{
		\begin{array}{l@{~}c@{~}l}
		  \fc & {\isot} & \frac{1}{\sqrt{2}} (\fc + \tc) \\
		  \tc & {\isot} & \frac{1}{\sqrt{2}} (\fc - \tc) \\
		\end{array}
	  \right\}
    \end{array}
    \qquad\qquad
    \begin{array}{ll}
      \opn{CNOT} : (\one\oplus\one)^{\tensor2}\isot (\one\oplus\one)^{\tensor2} \\
      \quad = \left\{
		\begin{array}{l@{~}c@{~}l}
		  \pv{\tc}{\fc} & {\isot} & \pv{\tc}{\tc} \\
		  \pv{\tc}{\tc} & {\isot} & \pv{\tc}{\fc} \\
		  \pv{\fc}{x}   & {\isot} & \pv{\fc}{x}
		\end{array}
  \right\} \end{array}\]
In this setting, any quantum circuit can be encoded by an iso.
\end{exa}

\subsection{Encoding into the Many-Worlds}

One can encode any term of the language into a Many-Worlds diagram.
Given some typing derivation $\xi$ of a term $x_1 : A_1, \dots, x_n :
A_n \vdash_t t : B$ we write  $\pparenthesis{\xi}$ for the function
that maps $\xi$ to a diagram in the Many-Worlds Calculus with $n$ input
wires of type $A_1, \dots, A_n$ and one output wire of type $B$.
For the typing derivation $\xi$ of an iso $\vdash_\omega \omega :A
\isot B$, $\pparenthesis{\xi}$ gives a diagram with one input wire of
type $A$ and one output wire of type $B$.\\
$\pparenthesis{\xi}$ is defined inductively over $\vdash_t$ and
$\vdash_\omega$ as shown in Figure~\ref{fig:translation}, where the worlds
sets are handled by world-agnostic compositions of diagrams with their
canonical labelings. This encoding can be shown to be sound in regard
to the programming language rewriting system~\cite{phd-kostia}.

\begin{exa}
\label{ex:iso-to-MW}
	We can represent the term
	\[t:=\left\{\begin{matrix} \langle\tc,x\rangle
	\leftrightarrow & \textbf{let }y=\textup{H }x\textup{ in
	}\frac{1}{\sqrt{2}}\langle \tc,y\rangle +
	\frac{1}{\sqrt{2}} \langle \fc,y\rangle\\
	\langle\fc,x\rangle \leftrightarrow & \textbf{let
	}y=\textup{Id }x\textup{ in }\frac{1}{\sqrt{2}}\langle
	\tc,y\rangle - \frac{1}{\sqrt{2}} \langle
	\fc,y\rangle \end{matrix} \right\}\] of type
	$(\one\oplus\one)^{\tensor2}\leftrightarrow(\one\oplus\one)^{\tensor2}$
	(already given in \cite{sabry2018symmetric}) as shown in
	Figure~\ref{fig:ex_match}. The yellow box stands for the Hadamard gate
	(which one can build following Example~\ref{ex:Hadamard-in-language}).
	Each line of this isomorphism corresponds to a column of the
	figure. Each columns start by matching the input as
	$\langle\tc,x\rangle$ or $ \langle\fc,x\rangle$,
	then computing $y$ from $x$, and finally building the output by following
	the syntax. The world set is computed by composing each of the
	blocks of this term, using the world-agnostic composition of
	$\ManyWorlds{\forall}$. It can be seen as a subset of $\{a,b\}
	\times \{c,c'\} \times \{0,1,2,3\}$ where $\{a,b\}$ corresponds to
	being on the first or second line of the matching, $\{c,c'\}$
	being on the left or right of the sum, and $\{0,1,2,3\}$ being the
	world set of the Hadamard gate.
\end{exa}

\begin{figure*}[!ht]
	\[
	\left\lParen{\begin{array}{c}\infer{x : A \vdash_e x :
			A}{}\end{array}}\right\rParen ~=~ \tikzfig{terms/var}
	\qquad\qquad
	\left\lParen{\begin{array}{c}\infer{\vdash_e \langle\rangle : \one}{}\end{array}}\right\rParen ~=~ \tikzfig{terms/unit}
	\]
	\[
	\left\lParen{\begin{array}{c}\infer{\Delta \vdash_e \omega~t :
		B}{\infer{\vdash_\omega \omega : A\isot B}{\xi_1} \qquad
		\infer{\Delta\vdash_e t : A}{\xi_2}}\end{array}}\right\rParen ~=~
		\tikzfig{terms/app}
	\]
	\[
	\left\lParen{\begin{array}{c}\infer{\vdash_e \isobasique : A
			\isot B}{\infer{\Delta_i \vdash_e v_i : A}{\xi_i}\qquad
			\infer{\Delta_i \vdash_e e_i :
				B}{\xi'_i}}\end{array}}\right\rParen ~=~\tikzfig{terms/iso}
	\]\vspace{-0.5cm}
	\[
	\left\lParen{\begin{array}{c}\infer{\Delta\vdash_e t_1 + t_2 :
			A}{\infer{\Delta\vdash_e t_1 : A}{\xi_1} \qquad
			\infer{\Delta\vdash_e t_2 : A}{\xi_2}}\end{array}} \right\rParen
	~=~ \tikzfig{terms/plus}
	\qquad\qquad
	\left\lParen{\begin{array}{c}\infer{\Delta\vdash_e\alpha t : A}{\infer{\Delta\vdash_e t : A}{\xi}}\end{array}}\right\rParen ~=~ \tikzfig{terms/alpha}
	\qquad\quad
	\]
	\[
	\left\lParen{\begin{array}{c}\infer{\Delta_1,\Delta_2\vdash_e \pv{t_1}{t_2} : A\otimes
		B}{\infer{\Delta_1\vdash_e t_1 : A}{\xi_1}\qquad \infer{\Delta_2\vdash_e
		t_2 : B}{\xi_2}}\end{array}}\right\rParen ~=~ \tikzfig{terms/tensor}
	\]
	\[
	\left\lParen{\begin{array}{c}\infer{\Delta_1, \Delta_2 \vdash_e \letpv{x_1, \dots, x_n}{t_1}{t_2} :
			B}{\infer{\Delta_1 \vdash_e t_1 : A_1\otimes \dots \otimes A_n}{\xi_1}
			\qquad\infer{x_1 : A_1, \dots, x_n : A_n, \Delta_2 \vdash_e t_2 :
				B}{\xi_2}}\end{array}}\right\rParen
	~=~ \tikzfig{terms/let}
	\]
	\[
	\left\lParen{\begin{array}{c}\infer{\Delta \vdash_e \inr{t} : A\oplus
			B}{\infer{\Delta \vdash_e t : A}{\xi}}\end{array}}\right\rParen ~=~ \tikzfig{terms/right}
	\qquad\qquad
	\left\lParen{\begin{array}{c}\infer{\Delta \vdash_e \inl{t} : A\oplus
			B}{\infer{\Delta \vdash_e t : B}{\xi}}\end{array}}\right\rParen ~=~ \tikzfig{terms/left}
	\]
	\caption{Translation of the Language}
	\label{fig:translation}
\end{figure*}

\section{Semantics of the Many-Worlds Calculus}
\label{sec:semantics}

Our calculus represents linear operators between finite dimensional
$R$-semimodules $(\FdM)$, or equivalently $R$-weighted matrices. In particular, in the case $R = \mathbb{C}$ we use linear
operators between finite dimensional Hilbert spaces, which correspond to
pure quantum computations. More precisely, we will define two
semantics, a world-dependent semantics $\interp{-}_a$ for every world
$a \in W$, which will be a monoidal functor from $\ManyWorlds{W}$ to
$\FdM$, and a world-agnostic semantics $\interp{-}$ from
$\ManyWorlds{\forall}$ to $\FdM$.

\subsection{Finite Dimensional \texorpdfstring{$R$}{R}-Semimodules} Similarly to $\mathbb{C}$-vector space, a $R$-semimodule is a set $M$ in which one can compute $R$-weighted sums of elements on $M$. More precisely, we have $+ : M \times M \to M$ and $0 \in M$ forming a semigroup, and $\cdot : R \times M \to M$ which is associative, left-distributive and right-distributive. Morphisms between $R$-semimodules are expected to preserve $R$-weighted sums (including the trivial sum $0$) that is:
	\[ f\left(\sum_{i=1}^n \lambda_i \cdot m_i\right) = \sum_{i=1}^n \lambda_i \cdot f(m_i)\]
	As in the case of vector spaces, it is said finite dimensional if there exists a finite basis, that is a finite set such that every element of $M$ can be uniquely decomposed as a $R$-weighted sum of that set. Relying on the uniqueness of this decomposition, a finite dimensional $R$-semimodule is actually isomorphic to $R^n$ for some $n \geq 0$. Note that our definition of finite dimensionality forces the semimodule to be \emph{freely} generated from a finite number of elements. Similarly to the vector space case, it is enough to define morphisms on a basis.
	For convenience, we will assume that $R$-semimodules come with a canonical basis, and for the remaining of this section we will write $\{m_1,\dots,m_m\}$ for the canonical basis of $M$ and $\{n_1,\dots,n_n\}$ for the canonical basis of $N$.
	This canonical basis allows us to define an operation $\langle~\mid~\rangle : M \times M \to R$, alike to the inner product\footnote{Contrary to the inner-product of Hilbert space, this operation does not take the conjugate of its left-hand-side parameter.} of vector spaces. Writing $\{m_1,\dots,m_m\}$ for the canonical basis of $M$, we define:
	\[ \left\langle \sum_{i=1}^m \lambda_i \cdot m_i ~~\middle|~~ \sum_{i=1}^m \rho_i \cdot m_i \right\rangle := \sum_{i=1}^m \lambda_i \times \rho_i \]
	While issues might arise in infinite dimensional cases, one can easily define the direct sum and the tensor product in our finite dimensional case. Indeed, we can define $M \oplus N$ and $M \otimes N$ as being the $R$-semimodules freely generated respectively by the pairs $(0,m_i)$ and $(1,n_j)$, and by the pairs $(m_i,n_j)$ which we write $m_i \otimes n_j$, that is:
\[ M \oplus N := \left\{ \sum_{1 \leq i \leq m} \lambda_{i} \cdot (0,m_i) + \sum_{1 \leq j \leq n} \rho_j \cdot (1,n_j) ~\middle|~ \lambda_{i},\rho_j \in R \right\} \]\[ M \otimes N := \left\{ \sum_{\substack{1 \leq i \leq m \\ 1 \leq j \leq n}} \lambda_{i,j} \cdot m_i \otimes n_j ~\middle|~ \lambda_{i,j} \in R \right\} \]
	Choosing a different basis for $M$ or $N$ yields an isomorphic $R$-semimodule, and this operation is associative and symmetric up to isomorphism. The $\otimes$ operation can be extended to a bilinear operation on elements of $M$ and $N$ as follows:
	\[ \left(\sum_{i=1}^m \lambda_i \cdot m_i\right) \otimes \left(\sum_{j=1}^n \rho_j \cdot n_j\right) := \sum_{i=1}^m\sum_{j=1}^n (\lambda_i \times \rho_j) \cdot m_i\otimes n_j \]
We write $\FdM(M,N)$ for morphisms between finite dimensional $R$-semimodules, and we note that it is itself a finite dimensional $R$-semimodule, meaning that we can compute $R$-weighted sums of morphisms. Indeed, the following is a basis of $\FdM(M,N)$:\[\left\{ m_i \mapsto n_j \text{ and } \forall k \neq i, m_k \mapsto 0 \right\}_{\substack{1 \leq i \leq m \\ 1 \leq j \leq n}} =  \left\{\sum_{k=1}^m \lambda_k \cdot m_k \mapsto \lambda_i \cdot n_j\right\}_{\substack{1 \leq i \leq m \\ 1 \leq j \leq n}}\]
This basis allows us to represent morphisms a $m \times n$ matrix with coefficients in $R$, and in fact composition of morphisms matches the usual matrix product. The operations $\cdot$, $+$ and $\otimes$ extend to morphisms, and correspond to multiplying every coefficient of the matrix by the same scalar, making a coefficient-by-coefficient sum, and making the Kronecker product of matrices. Taking inspiration from matrices, a transpose operation can be defined as follows. If $f \in \FdM(M,N)$ satisfies $f(m_i) = \sum_{j=1}^n \lambda^i_j n_j$, then $f^t \in \FdM(N,M)$ is defined as:
\[ f^t\left(\sum_j \rho_j \cdot n_j\right) := \sum_{\substack{1 \leq i \leq m \\ 1 \leq j \leq n}} \rho_j \times \lambda^i_j \cdot m_i \]
We also note that block matrices work as expected: a morphism $f \in \FdM(M_0 \oplus M_1,N_0 \oplus N_1)$ can be seen as four morphisms $f_{ij} \in \FdM(M_i,N_j)$ assembled together as $f = \begin{pmatrix}
f_{00} & f_{10} \\ f_{01} & f_{11}
\end{pmatrix}$, that is, writing $m_k^i$ and $n_k^j$ for the elements of the canonical basis of $M_i$ and $N_j$ respectively, and writing $f_{ij}(m_k^i) = \sum_{\ell} \lambda_{\ell}^{i,j,k} \cdot n_k^j$ then we have:
\[ f\left(\sum_{i \in \{0,1\}} \sum_k \rho_k^i \cdot (i,m_k^i) \right) = \sum_ {i,j \in \{0,1\}} \sum_{k,\ell} \rho_k^i \times \lambda_{\ell}^{i,j,k} \cdot (j,n_k^j) \]
From the above properties it follows that $\FdM$ forms a category, with $\otimes$ and $\oplus$ being two symmetric monoidal products, $\otimes$ distributive over $\oplus$ and $\oplus$ being a biproduct.

\begin{figure*}[!ht]
	\begin{align*}
	\interp{\tikzfig{explicit/swap-w}}_a &= \left\lbrace
	\begin{array}{lll}
	\textup{Id}
	&\in \FdM(\mathcal{M}_{A},\mathcal{M}_A) &\text{ if }a \in w\backslash v \\
	\textup{Id}
	&\in \FdM(\mathcal{M}_{B},\mathcal{M}_B) &\text{ if }a \in v\backslash w \\
	h \otimes h' \mapsto h' \otimes h &\in \FdM(\mathcal{M}_{A\tensor B},\mathcal{M}_{B \tensor A}) &\text{ if }a \in w \cap v\\
	(1) &\in \FdM(R,R) &\text{ otherwise} \end{array}\right.\\
	\interp{\tikzfig{explicit/cup-w}}_a &= \left\lbrace
	\begin{array}{lll}
	h \otimes h' \mapsto \langle h | h' \rangle &\in \FdM(\mathcal{M}_{A\tensor A},R) &\text{ if }a \in w \\
	(1) &\in \FdM(R,R) &\text{ otherwise} \end{array}\right.\\
	\interp{\tikzfig{explicit/contraction-w}}_a &= \left\lbrace
	\begin{array}{lll}
	\textup{Id}  &\in \FdM(\mathcal{M}_{A},\mathcal{M}_A) &\text{ if }a \in \bigsqcup_i w_i \\ (1) &\in \FdM(R,R) &\text{
		otherwise} \end{array}\right. \\
	\interp{\tikzfig{explicit/scalar-w}}_a &= \left\lbrace
	\begin{array}{lll}
	s \cdot \textup{Id} &\in \FdM(\mathcal{M}_A,\mathcal{M}_A) &\text{ if }a \in w \\
	(1) &\in \FdM(R,R) &\text{ otherwise} \end{array}\right. \\
	\interp{\tikzfig{explicit/plus-w}}_a &= \left\lbrace
	\begin{array}{lll} \begin{pmatrix}
	\textup{Id} \\ 0
	\end{pmatrix} &\in \FdM(\mathcal{M}_A,\mathcal{M}_{A\oplus B}) &\text{ if }a \in w \\
	\begin{pmatrix}
	0 \\ \textup{Id} \end{pmatrix} &\in \FdM(\mathcal{M}_B,\mathcal{M}_{A\oplus
		B}) &\text{ if }a \in v\\ (1) &\in \FdM(R,R) &\text{
		otherwise} \end{array}\right.\\
	\interp{\tikzfig{explicit/tensor-unit-w}}_a &= \left\lbrace
	\begin{array}{lll}
	(1)  &\in \FdM(\mathcal{M}_{\varnothing},\mathcal{M}_{\one}) &\text{ if }a \in w \\ (1) &\in \FdM(R,R) &\text{
		otherwise} \end{array}\right. \\
	\interp{\tikzfig{explicit/tensor-PN-w}}_a &= \left\lbrace
	\begin{array}{lll}
	\textup{Id}  &\in \FdM(\mathcal{M}_{A} \otimes \mathcal{M}_{B},\mathcal{M}_{A \tensor B}) &\text{ if }a \in w \\ (1) &\in \FdM(R,R) &\text{
		otherwise} \end{array}\right.\\
	\end{align*}
	\caption{Semantics of the Generators of $\ManyWorlds{W}$ in a
		World $a \in W$. Mirrored generators use the transposed for
		their semantics.}
	\label{fig:semantics_gen}
\end{figure*}
\subsection{Semantics of Objects}

We start by defining the semantics of $\ManyWorlds{W}$ and $\ManyWorlds{\forall}$ on the objects. For every object
$\mathcal{A}$ of \catD, we define its \emph{enablings} $\enab{\mathcal{A}}$
as ``replacing any number of wire types by $\bullet$''. For example
$\enab{(A \parallel B)} = \{ \bullet \parallel \bullet, \bullet
\parallel B, A \parallel \bullet, A \parallel B \}$.
Then, for any object $(\mathcal{A},\ell_{\mathcal{A}})$ of $\ManyWorlds{W}$
and any world $a \in W$, we define
$\enabW{(\mathcal{A},\ell_{\mathcal{A}})}{a}$ to be the enabling of
$\mathcal{A}$ in which every $(A:w)$ with $a \in w$ is preserved and
every $(A:w)$ with $a \notin w$ is replaced by $\bullet$. For example
$\enabW{(A:\{a\}\parallel B:\{b\})}{a} = A \parallel \bullet$.
To each enabling
$\mathcal{E} \in \enab{\mathcal{A}}$ we associate an $R$-semimodule
$\mathcal{M}_{\mathcal{E}}$ as follows:
\[
\begin{matrix}
\mathcal{M}_{A_1\parallel \dots \parallel A_n} := \mathcal{M}_{A_1}
\otimes \dots \otimes \mathcal{M}_{A_n} \\
\mathcal{M}_{\emptysquare} := R \qquad  \mathcal{M}_{\bullet}
:= R \qquad \mathcal{M}_{\one} := R \qquad \mathcal{M}_{\zero} := \set{0} \\
\mathcal{M}_{A \tensor B} := \mathcal{M}_{{A}} \otimes
\mathcal{M}_{{B}}
\qquad
\mathcal{M}_{A \oplus B} := \mathcal{M}_{{A}} \oplus \mathcal{M}_{{B}}
\end{matrix}
\]
We can then define the semantics $\interp{-}_a:\ManyWorlds{W} \to
\FdM$ and $\interp{-}:\ManyWorlds{\forall} \to \FdM$ on objects as
$\interp{(\mathcal{A},\ell_{\mathcal{A}})}_a :=
\mathcal{M}_{\enabW{(\mathcal{A},\ell_{\mathcal{A}})}{a}}$ and
$\interp{\mathcal{A}} := \bigoplus_{\mathcal{E} \in
\enab{\mathcal{A}}} \mathcal{M}_{\mathcal{E}}$.

\subsection{World-Dependent Semantics}

Then, for the morphisms, we proceed by compositionality for
$\interp{-}_a$, meaning that we define $\interp{-}_a$ on every
generator and compute the semantics of a diagram by decomposing it
with
$\interp{g \circ f}_a := \interp{g}_a \circ \interp{f}_a$ and
$\interp{f \parallel g}_a := \interp{f}_a \otimes \interp{g}_a$.
The semantics of all the generators is given in
Figure~\ref{fig:semantics_gen}.

\subsection{World-Agnostic Semantics}
The world-agnostic semantics is defined from the world-dependent
semantics, as follows. Consider $f$ a morphism of
$\ManyWorlds{W}((\mathcal{A},\ell_{\mathcal{A}}),(\mathcal{B},\ell_{\mathcal{B}}))$,
we define its world-agnostic semantics $\interp{\withWorld{f}{W}}\in \FdM\left(
\bigoplus_{\mathcal{E} \in \mathcal{A}^{\bullet}}
\mathcal{M}_{\mathcal{E}}, \bigoplus_{\mathcal{F} \in
\mathcal{B}^{\bullet}} \mathcal{M}_{\mathcal{B}} \right)$ as:

\[\interp{\withWorld{f}{W}} := \left\{
		\text{\raisebox{15pt}{$\sum_{\substack{a\in W
		\\\enabW{(\mathcal{A},\ell_{\mathcal{A}})}{a}=\mathcal{A}'\\\enabW{(\mathcal{B},\ell_{\mathcal{B}})}{a}=\mathcal{B}'}}
		\interp{f}_a$}}\right\}_{\mathcal{A}' \in
		\mathcal{A}^{\bullet}, \mathcal{B}' \in \mathcal{B}^{\bullet}}
		\]

For example, the worlds-agnostic semantics of the generators (see
Figure~\ref{fig:world_agnostic_semantics_gen}) are simply the collection of
all their world-dependent semantics assembled into a single linear
operator, as one can see by comparing Figure~\ref{fig:semantics_gen}
to Figure~\ref{fig:world_agnostic_semantics_gen}. We first show that the
worlds-agnostic semantics is functorial:

\begin{figure*}[t]
\begin{align*}
\interp{\tikzfig{explicit/swap}} &=\begin{blockarray}{@{}ccccc@{}}
& \overset{A \parallel B}~ & \overset{A \parallel \bullet}~ & \overset{\bullet \parallel B}~ & \overset{\bullet \parallel \bullet}~ \\
\begin{block}{c(cccc)}
\scalebox{0.8}{$B \parallel A$}& \scalebox{0.7}{$\begin{bmatrix} h \otimes h'\\ \mapsto\\ h' \otimes h \end{bmatrix}$}  & 0 & 0 & 0 \\
\scalebox{0.8}{$B \parallel \bullet$}& 0 & 0 & \textup{Id} & 0 \\
\scalebox{0.8}{$\bullet \parallel A$}& 0 & \textup{Id} & 0 & 0 \\
\scalebox{0.8}{$\bullet \parallel \bullet $} & 0 & 0 & 0 & 1 \\
\end{block}
\end{blockarray}&&\\
\interp{\tikzfig{explicit/cup}} &=\begin{blockarray}{ccccc}
& \overset{A \parallel A}~ & \overset{A \parallel \bullet}~ & \overset{\bullet \parallel A}~ & \overset{\bullet \parallel \bullet}~ \\
\begin{block}{c(cccc)}
\scalebox{0.8}{$\bullet$}& \scalebox{0.7}{$\begin{bmatrix} h \otimes h'\\ \mapsto\\ \langle h | h' \rangle \end{bmatrix}$}  & 0 & 0 & 1 \\
\end{block}
\end{blockarray}&&\\
\interp{\tikzfig{explicit/scalar}} &=\begin{blockarray}{ccc}
& \overset{A}~ & \overset{\bullet}~\\
\begin{block}{c(cc)}
\scalebox{0.8}{$A$}& s \cdot \textup{Id} & 0 \\
\scalebox{0.8}{$\bullet$} & 0 & 1 \\
\end{block}
\end{blockarray}&&\\
\interp{\tikzfig{explicit/plus}} &=
\begin{blockarray}{ccccc}
& \overset{A \parallel B}~ & \overset{A \parallel \bullet}~ & \overset{\bullet \parallel B}~ & \overset{\bullet \parallel \bullet}~ \\
\begin{block}{c(cccc)}
\!\scalebox{0.8}{$A \oplus B$}\!& \substack{0\\0} & \substack{\textup{Id}\\0} & \substack{0\\\textup{Id}} & \substack{0\\0} \\
\scalebox{0.8}{$\bullet$} & 0 & 0 & 0 & 1 \\
\end{block}
\end{blockarray}&&\\
\interp{\tikzfig{explicit/tensor-PN}} &=\begin{blockarray}{ccccc}
& \overset{A \parallel B}~ & \overset{A \parallel \bullet}~ & \overset{\bullet \parallel B}~ & \overset{\bullet \parallel \bullet}~ \\
\begin{block}{c(cccc)}
\scalebox{0.8}{$A \tensor B$}& \textup{Id} & 0 & 0 & 0 \\
\scalebox{0.8}{$\bullet$} & 0 & 0 & 0 & 1 \\
\end{block}
\end{blockarray}&
\interp{\tikzfig{explicit/tensor-unit}} &=\begin{blockarray}{cc}
& \overset{\bullet}~ \\
\begin{block}{c(c)}
\scalebox{0.8}{$\one$}& \textup{Id}\\
\scalebox{0.8}{$\bullet$} & 1 \\
\end{block}
\end{blockarray}\\
\interp{\tikzfig{explicit/contraction_bin}} &=\begin{blockarray}{c cccc}
& \overset{A \parallel A}~ & \overset{A \parallel \bullet}~ & \overset{\bullet \parallel A}~ &\overset{\bullet \parallel \bullet}~ \\
\begin{block}{c(cccc)}
\scalebox{0.8}{$A$}& 0 & \textup{Id} & \textup{Id} & 0\\
\scalebox{0.8}{$\bullet$} & 0  & 0 & 0 & 1\\
\end{block}
\end{blockarray}&
\interp{\tikzfig{explicit/contraction_null}} &=\begin{blockarray}{c c}
&\overset{\bullet}~ \\
\begin{block}{c(c)}
\scalebox{0.8}{$A$}&  0\\
\scalebox{0.8}{$\bullet$} & 1\\
\end{block}
\end{blockarray}
\end{align*}
Or more generally:
\[
\interp{\tikzfig{explicit/contraction}} =\begin{blockarray}{c cccccc}
& \overset{A \parallel \dots \parallel A}~ & \overset{\dots}~ & \overset{A \parallel \bullet \parallel \dots \parallel \bullet}~ & \overset{\dots}~ & \overset{ \bullet \parallel \dots \parallel \bullet \parallel A}~ &\overset{\bullet \parallel \dots \parallel \bullet}~ \\
\begin{block}{c(cccccc)}
\scalebox{0.8}{$A$}& 0 & 0 & \textup{Id} & \textup{Id} & \textup{Id} & 0\\
\scalebox{0.8}{$\bullet$} & 0 & 0 & 0 & 0 & 0 & 1\\
\end{block}
\end{blockarray} \]

\caption{World Agnostic Semantics of the Generators.\\ Mirrored generators use the transposed for
their semantics.}
\label{fig:world_agnostic_semantics_gen}
\end{figure*}

\begin{prop}
	\label{prop:world-agnostic-semantics-functorial}
	The worlds-agnostic semantics $\interp{-}$ defined in
	Section~\ref{sec:semantics} is a monoidal functor from
	$\ManyWorlds{\forall}$ to $\FdM$.
\end{prop}
\begin{proof}
	We recall here the definition of the worlds-agnostic semantics of
	$\withWorld{f}{W} : \mathcal{A} \to \mathcal{B}$:
	\[\interp{\withWorld{f}{W}} := \left\{
	\text{\raisebox{15pt}{$\sum_{\substack{a\in W
	\\\enabW{(\mathcal{A},\ell^f_{\mathcal{A}})}{a}=\mathcal{A}'\\\enabW{(\mathcal{B},\ell^f_{\mathcal{B}})}{a}=\mathcal{B}'}}
	\interp{f}_a$}}\right\}_{\mathcal{A}' \in \mathcal{A}^{\bullet},
	\mathcal{B}' \in \mathcal{B}^{\bullet}}
	\]

	From the definition of the worlds-agnostic compositions, we
	directly have:
	\[\interp{\withWorld{f}{W} \parallel \withWorld{g}{V}}_{(a,b)} = \interp{\withWorld{f}{W}}_a
	\otimes \interp{\withWorld{g}{V}}_b \]
	\[\text{and}\qquad\interp{\withWorld{g}{V} \circ
	\withWorld{f}{W}}_{(a,b)} = \interp{\withWorld{g}{V}}_b \circ
	\interp{\withWorld{f}{W}}_a\] Remember that in the first case the
	set of worlds is $W \times V$, while in the second case, it is
	included in it. The functoriality with respect to the parallel
	composition is then immediate:
	\begin{align*}
	&\interp{\withWorld{f}{W} \parallel \withWorld{g}{V}}
	= \left\{ \sum \interp{\withWorld{f}{W} \parallel \withWorld{g}{V}}_{(a,b)} \right\}
	= \left\{ \sum \interp{\withWorld{f}{W}}_a\right\} \otimes  \left\{\sum \interp{\withWorld{g}{V}}_{b} \right\}
	= \interp{\withWorld{f}{W}} \otimes \interp{\withWorld{g}{V}}
	\end{align*}
	The functoriality with respect to the sequential composition is
	more subtle, as one must carefully manipulate the indices of the
	sum and remark that the set of worlds $w$ eliminated by the
	worlds-agnostic composition satisfies the following:
	\[ (a,b) \notin w \iff
	\enabW{(\mathcal{B},\ell^f_{\mathcal{B}})}{a} =
	\enabW{(\mathcal{B},\ell^g_{\mathcal{B}})}{b}\] where $f :
	(\mathcal{A},\ell^f_{\mathcal{A}}) \to
	(\mathcal{B},\ell^f_{\mathcal{B}})$ and $g :
	(\mathcal{B},\ell^g_{\mathcal{C}}) \to
	(\mathcal{C},\ell^g_{\mathcal{C}})$.

	Then, we have
	\begin{align*}
	&\interp{ \withWorld{g}{V} \circ \withWorld{f}{W}}
	= \left\{ \sum \interp{\withWorld{g}{V} \circ \withWorld{f}{W}}_{(a,b)} \right\}= \left\{\sum \interp{\withWorld{g}{V}}_{b} \right\} \circ \left\{ \sum \interp{\withWorld{f}{W}}_a\right\}
	= \interp{\withWorld{g}{V}} \circ \interp{\withWorld{f}{W}} \qedhere
\end{align*}
 \end{proof}

In order to show the universality of our language, we first start by
defining an equational theory, that we show is sound, then define a
notion of normal form. Furthermore, we will show that the normal form
is unique, which is needed to prove the completeness of the language.

\section{The Equational Theory}
\label{sec:equation}
\begin{figure*}[!ht]
	\centering
	\scalebox{.95}{$\tikzfig{eq/all}$}
	\caption{Equations with a Fixed World Set $W$}
	\label{fig:eq_incomplete}
\end{figure*}

\begin{figure*}[!htb]
\vspace*{3em}
	\[\tikzfig{eq/all_complete}\]
	\caption{Equations with Side-Effects on World Sets}
	\label{fig:eq_complete}
\end{figure*}

Similarly to how our semantics is defined in two steps, the
equational theory is also defined in two steps:
\begin{enumerate}
\item
A set of equations within $\ManyWorlds{W}$ for a fixed set of worlds
$W$, which will not be complete, but will be sound for $\interp{-}_a$
for every $a \in W$, hence sound for $\interp{-}$ too. We write
$\withWorld{\equiv}{W}$ for the induced congruence\footnote{In other
words the smallest equivalence relation satisfying those equations and
such that $f \withWorld{\equiv}{W} f' \implies \forall g,h,l, g \circ
(f \parallel h) \circ k \withWorld{\equiv}{W} g \circ (f' \parallel h)
\circ k$.} over $\ManyWorlds{W}$. We list those equations in
Figure~\ref{fig:eq_incomplete}. For all but one of those equations
that is restricted to the type $\zero$, they can be applied for wires
of any type. Quite notably, the last two rows describe the fact that
the contraction is a natural transformation.

\item
Five additional equations with side effects on the set of worlds,
which will be sound and complete for $\interp{-}$ (but not for
$\interp{-}_a$). We write $\equiv$ for the induced equivalence
relation, which is a congruence over $\ManyWorlds{\forall}$.  We have:
One equation that allows us to rename the worlds: for every morphism
$(\D_f,\ell_f)$ of $\ManyWorlds{W}$, and for every bijection $i : W
\to V$, we have $\withWorld{(\D_f,\ell_f)}{W} \equiv \withWorld{(\D_f,i \circ \ell_f)}{V}$;
Two equations allowing the annihilation (or creation, when looking at
them from right to left) of worlds due to coproducts or scalars (first
row of Figure~\ref{fig:eq_complete}); Two equations allowing the splitting
(or merging, when looking at them from right to left) of worlds due to
coproducts or scalars (second row of Figure~\ref{fig:eq_complete}).
\end{enumerate}

\begin{prop}[Soundness]\label{prop:soundness}
	For $f$ a morphism of $\ManyWorlds{W}$ and $g$ of
        $\ManyWorlds{V}$, whenever $\withWorld{f}{W} \equiv \withWorld{g}{V}$ we have $\interp{\withWorld{f}{W}} =
        \interp{\withWorld{g}{V}}$. Additionally if $W = V$, whenever $f \withWorld{\equiv}{W} g$
        we have $\forall a \in W, \interp{f}_a = \interp{g}_a$.
\end{prop}
\begin{proof}

Given that most of the time, $\interp{-}_a$ is the
identity, the equations defining $\withWorld{\equiv}{W}$ are quite straightforward
to verify. We immediately have that $\withWorld{\equiv}{W}$ is sound with respect
to $\interp{-}_a$ for every $a \in W$. Since $\interp{-}$ is defined
from $\interp{-}_a$, soundness with respect to $\interp{-}$ is also
correct. We then handle the five additional equations of $\equiv$.

\noindent\textbf{Renaming.} Applying a bijection to the world set $W$
does not change the result computed by $\sum_{a \in W} \dots$, hence
this equation is sound with respect to $\interp{-}$.

\noindent\textbf{Annihilation due to Scalars.} This equation simply
removes elements equal to zero from the sum $\sum_{a \in W} \dots$,
hence it is sound with respect to $\interp{-}$.

\noindent\textbf{Annihilation due to Plus.} Since $\oplus$ is a
	 biproduct in $\FdM$, we have $\textup{proj}_H^{H
	 \oplus K} \circ \textup{inj}_H^{H \oplus K} = \id_{H}$,
	 $\textup{proj}_K^{H \oplus K} \circ \textup{inj}_K^{H \oplus K} =
	 \id_{K}$, $\textup{proj}_K^{H \oplus K} \circ \textup{inj}_H^{H
	 \oplus K} = 0$ and $\textup{proj}_H^{H \oplus K} \circ
	 \textup{inj}_K^{H \oplus K} = 0$. One can then simply remove from
	 the $\sum_{a \in W} \dots$ the elements equal to zero, which
	 proves that \emph{Annihilation due to Plus} is sound with respect
	 to $\interp{-}$.

\noindent\textbf{Splitting due to Scalars.} Since $\FdM(R, S)$
is a $R$-semimodule, we have $(s+t)\cdot f = s \cdot f + t\cdot f$, which is exactly
the property required for this equation to be sound for $\interp{-}$.

\noindent\textbf{Splitting due to  Plus.} Similarly, we have in
$\FdM$ the property that $\id_{H \oplus K} =
\textup{inj}_H^{H \oplus K} \circ \textup{proj}_H^{H \oplus K} +
\textup{inj}_K^{H \oplus K} \circ \textup{proj}_K^{H \oplus K}$, which
is the property required for this equation to be sound for
$\interp{-}$.
\end{proof}

\begin{exa}[The Quantum Switch]\label{exa:qs}
	The Quantum Switch is a prime example of a process that makes use
	of quantum control flows. It assumes two operators $U$ and $V$ that can be
	applied as black boxes to a target system, and a control qubit that determines
	the order of application of these two operators. While the most direct
	implementation (e.g.~using the language defined in Remark~\ref{rem:star_worlds})
	makes two calls to the operators, one for each branch following the semantics:
	\[(\alpha\ket 0+\beta\ket1)\otimes \ket\Psi \mapsto
	\alpha\ket 0\otimes U\circ V\ket\Psi+\beta\ket1\otimes V\circ U\ket\Psi,\]
	physical implementations exist that only require one occurrence of each of
	the two operators.

	The leftmost diagram in Figure~\ref{fig:rewrite_QS} represents, within the Many-%
	Worlds calculus, the switch with a single occurrence of the operators,
	while the rightmost diagram uses the more direct, branch-wise representation.
	The equational theory offers the possibility to formally jump from one
	perspective to the other (as shown in the figure).

  In those diagrams, the world set is $W = w \sqcup v$ and we rely on
  violet, blue and red wires to indicate respectively worlds labels
  $w \sqcup v$, $w$ and $v$. Each figure has a control side which
  operates on a quantum bit (type $\one \oplus \one$) and binds the
  world $w$ to $\ket{0}$ and the world $v$ to $\ket{1}$; and a
  computational side which operates on some data of an arbitrary type
  $A$, on which could be applied $U$ and/or $V$ which stand for two
  morphisms of $\ManyWorlds{W}(A:W,A:W)$.

  \noindent
    The first rewriting step relies on the two following lemmas:
    $$\tikzfig{explicit/lemma_nat} \qquad\quad \tikzfig{explicit/lemma_c}$$
    both of which being deducible from the equational theory (see
    Section~\ref{sec:induced_eq}). The second rewriting step is simply using
    the properties of a compact closed category.
\end{exa}

\begin{figure*}[!ht]
  \[ \tikzfig{explicit/legend}   \qquad \tikzfig{explicit/QS_unique}
  \quad \equiv_{w \sqcup v} \quad \tikzfig{explicit/QS_middle} \quad
  \equiv_{w \sqcup v} \quad \tikzfig{explicit/QS_double} \]
  \caption{Rewriting the Quantum Switch}
  \label{fig:rewrite_QS}
\end{figure*}

To emphasise the fact that each black box is queried once, the
quantum switch is often presented as a higher-order operation where
$U$ and $V$ appear as arguments. Using Remark~\ref{rem:currying} we
can modify the leftmost diagram to represent the switch as a higher
order operation, as shown in Figure~\ref{fig:QS_higher}.

\begin{figure*}[!ht]
	\[\tikzfig{explicit/QS_higher} \]
	\caption{Higher Order Quantum Switch}
	\label{fig:QS_higher}
\end{figure*}

\section{Induced Equations}
\label{sec:induced_eq}

In Figure~\ref{fig:eq_incomplete}, we presented a set of equation reasonably
small, by having equations parameterized by the arity of the
contraction, and by omitting a lot of useful equations that can be
deduced from them. In Figure~\ref{fig:eq_alt}, we take the opposite
approach: we give equations using the contractions of arity zero and
two (which are sufficient to generate all the other contractions) and
we provide additional axioms that follows from
Lemma~\ref{lem:simple_induced_eq} and Lemma~\ref{lem:induced_nat}.

\begin{figure*}[!ht]
	\scalebox{0.9}{\tikzfig{eq/all_induced}}
	\caption{Alternative Presentation of the Equational Theory with a Fixed World Set $W$}
	\label{fig:eq_alt}
\end{figure*}

Note that this is only an alternative presentation to the equational
theory for $\withWorld{\equiv}{W}$, the axioms of Figure~\ref{fig:eq_complete} are still
required for $\equiv$.

\begin{lem}
	\label{lem:simple_induced_eq}
	Whenever $w_i$ are disjoints sets of worlds, we have the following:
	\[ \hfill \tikzfig{eq/c-cancel} \hfill \]
	\[\hfill \tikzfig{eq/scalar_jump} \hfill\]
	\[\hfill \tikzfig{eq/scalar_dup}\hfill \tag*{\qed}\]
\end{lem}

\begin{lem}
	\label{lem:induced_nat}
	Whenever $w_i$ and $v_j$ are disjoint sets of worlds, we have the following:
	\begin{itemize}
	\setlength\itemsep{1em}
	\item \tikzfig{eq/nat-induced-unit}
	\item \tikzfig{eq/nat-induced-tensor}
	\item \tikzfig{eq/nat-induced-plus}
	\end{itemize}
\end{lem}
\begin{proof}
	We provide a proof for the third equation, the first two are proven similarly.
	\[\tikzfig{eq/nat_induced_proof} \qedhere\]
\end{proof}

\begin{cor}[Naturality of the Binary Contraction]
\label{lem:contraction-natural}
	For every $f : \parallel^n (A_i:w_i) \to \parallel^m (B_j:v_j)$
	with world set $W$ and every $u \subseteq W$, we have
	\[\tikzfig{eq/nat_gen_bin}\]
	where $f\backslash u :  \parallel^n (A_i:w_i\backslash u) \to
	\parallel^m (B_j:v_j\backslash u)$ is equal to $f$ where every
	worlds label $w$ has been replaced by $w\backslash u$, and
	similarly for $f \cap w$.
\end{cor}
\begin{proof}
	This is proven by induction over $f$. All the generator cases
	(including Cup and Cap) follow directly from the equations given
	in Figure~\ref{fig:eq_incomplete} and Lemma~\ref{lem:induced_nat} together
	with the properties of a compact close category.
\end{proof}

\begin{lem}[Empty World]
	\label{lem:contraction-zero}
	For every $f : \parallel^n (A_i:\varnothing) \to \parallel^m
	(B_j:\varnothing)$ with world set $W$ but such that every worlds
	label of $f$ is $\varnothing$, we have
	\[ \tikzfig{eq/nat_gen_zero} \]
\end{lem}
\begin{proof}
	This is proven by replacing every wire by two contractions of
	arity zero (sixth axiom of Figure~\ref{fig:eq_incomplete} with $n = 0$), and
	then using the naturality of the contraction of arity zero (last two
	lines of Figure~\ref{fig:eq_incomplete} with $n = 0$) to consume every
	generator.
 \end{proof}

\subsection{Normal Form}
\label{sec:normal}
For the remaining of this section, we use $\equiv$ instead of $\withWorld{\equiv}{W}$, as we will occasionally use equations from Figure~\ref{fig:eq_complete}.
We are now ready to define the normal form of our diagrams, for that,
it will be practical to define the following syntactic sugar, which we
call the \emph{unitor}, its \emph{unit} and its \emph{generalized form}:
$$\tikzfig{unitor}~:=~\tikzfig{unitor-def}
\qquad\qquad\tikzfig{unit-of-unitor}
\qquad\qquad\tikzfig{generalized-unitor}~:=~\tikzfig{generalized-unitor-def}$$

 We define the following short-hand:
 \[\tikzfig{NF-matrix-block-gen}\]
  with the assumption that the
 world set is in bijection with the set of scalars, in other words
 the scalars $\lambda_{i\,j}$ have for worlds label
 $\{a_{i\,j}\}$. In
 particular, all the input (resp.~output) wires live in mutually
 exclusive worlds.

 An important observation is that any permutation
 of wires (all mutually exclusive, and of type $\one$) can easily be
 put in this form using the following equations:
 \[
   \phantom{.}\hfill\tikzfig{NF-dangling-scalar-0}\hfill\tikzfig{1-1-contraction-is-identity} \hfill
   \tikzfig{NF-0-branch}\hfill\phantom{.}
 \]

 \begin{defi}[Normal Form]
 The normal
 form of a morphism $f:\mathcal{A}\to \mathcal{B}$ is defined as follows:
 \[f = \tikzfig{NF-gen}\]

 where the morphisms $\iso_{\mathcal{A}}$ and $\iso_{\mathcal
 B}^{\inv}$ are defined inductively as:
 \[\tikzfig{iso-I-gen} \qquad\qquad \tikzfig{iso-1} \qquad\qquad \tikzfig{iso-0}\]
 \[\tikzfig{iso-plus-gen} \qquad\qquad \tikzfig{iso-tensor-gen}\]
 \[\tikzfig{iso-boxempty}\]

\end{defi}

The output wires of $\iso_{\mathcal{A}}$ for any $\mathcal{A}$ live in
mutually exclusive worlds, but once again, we don't overload the
diagrams with unitors or world names encoding this information,
although it will be used in the following.

Notice that graphically, there is no difference between $\mathcal
A\parallel(\mathcal B\parallel \mathcal C)$ and $(\mathcal A\parallel \mathcal
B)\parallel \mathcal C$, in other words $\parallel$ is strictly
associative. However $\iso_{\mathcal A\parallel(\mathcal B\parallel \mathcal
  C)}$ and $\iso_{(\mathcal A\parallel \mathcal B)\parallel \mathcal C}$ are
different, but they are equivalent up to a rearranging of the output
wires:
\begin{lem}
  \label{lem:quasi-associativity-of-parallel}
  There exists a wire permutation $\sigma$ such that $\iso_{\mathcal
  A\parallel(\mathcal B\parallel \mathcal C)}=\sigma\circ\iso_{(\mathcal
  A\parallel \mathcal B)\parallel\mathcal C}$.
  \qed
\end{lem}

\begin{proof}
\phantomsection\label{prf:quasi-associativity-of-parallel} First
notice that in both $\iso_{\mathcal A\parallel(\mathcal B\parallel \mathcal
C)}$ and $\iso_{(\mathcal A\parallel \mathcal B)\parallel \mathcal C}$, we can
use the bialgebra between contractions and unitors, followed by their
respective fusions in the following way:
\def\fig{iso-boxempty-associativity}
\begin{align*}
\InputIfFileExists{./figures/\fig/\fig_00.tikz}{}{{\color{red}\colorbox{pink}{missing file : 00}}}
&\equiv\InputIfFileExists{./figures/\fig/\fig_01.tikz}{}{{\color{red}\colorbox{pink}{missing file : 01}}}
\equiv\InputIfFileExists{./figures/\fig/\fig_02.tikz}{}{{\color{red}\colorbox{pink}{missing file : 02}}}\\
&\equiv\InputIfFileExists{./figures/\fig/\fig_03.tikz}{}{{\color{red}\colorbox{pink}{missing file : 03}}}
\end{align*}
and similarly for $\iso_{(A\parallel B)\parallel C}$. It then suffices
to check which contractions the bottom unitors are linked to. Naming
the $i$-th contraction exiting $\iso_A$ as $a_i$, and
similarly for $B$ and $C$, we can see that for each triple
$(a_i,b_j,c_k)$ there is exactly one unitor connected to precisely
contraction $a_i$, $b_j$ and $c_k$, in both diagrams. The same is true
for every pair $(a_i,b_j)$, $(a_i,c_k)$ and $(b_j,c_k)$, as well as
for every 1-tuple $(a_i,)$, $(b_j,)$ and $(c_k,)$. This shows that
both diagrams are equal up to rearranging of the outputs.
\end{proof}

We hence have a choice to make here for canonicity, and choose
$\iso_{A_0\parallel A_1\parallel A_2\parallel
  ...}:=\iso_{(...((A_0\parallel A_1)\parallel A_2)\parallel ...)}$.

We define $\iso_{\mathcal{A}}^{\inv}$ inductively in the same way, but
upside-down.
We note that $\iso_{\mathcal{A}}^{-1}\circ\iso_{\mathcal{A}}$ is the
normal form of $\textup{id}_{\mathcal{A}}$.
\begin{lem}
Notice that \tikzfig{iso-plus-wrt-boxempty} and
\tikzfig{iso-tensor-wrt-boxempty}.
\end{lem}

\begin{lem}
	\label{lem:isos-are-isos}
	We have the following identities:
	\[\tikzfig{isos-are-isos}~~\text{ and }~~\tikzfig{ok-isos-are-actually-pseudo-isos}\]
	\text{where $\ell(s_i)\cap\ell(s_j)=\varnothing$ when $i\neq j$.}
\end{lem}

\begin{proof}
\label{prf:isos-are-isos}
We will use the following identities:
\begin{align}
\tikzfig{isos-are-isos-boxempty}
\end{align}

when all the wires  $a_{ij}, b_i,c_j$ are mutually exclusive.

We can show this result by induction on $n$ and $m$. Case $(0,m)$ is
obvious. Case $(1,1)$ can be proven easily using world sets:
\def\fig{isos-are-isos-boxempty-prf-1-1-case}
\begin{align*}
\InputIfFileExists{./figures/\fig/\fig_00.tikz}{}{{\color{red}\colorbox{pink}{missing file : 00}}}
~~\equiv~~\InputIfFileExists{./figures/\fig/\fig_01.tikz}{}{{\color{red}\colorbox{pink}{missing file : 01}}}
~~\equiv~~\InputIfFileExists{./figures/\fig/\fig_02.tikz}{}{{\color{red}\colorbox{pink}{missing file : 02}}}
~~\equiv~~\InputIfFileExists{./figures/\fig/\fig_03.tikz}{}{{\color{red}\colorbox{pink}{missing file : 03}}}
\end{align*}
For any $n$ and $m$, we can then prove the case $(n+1,m)$ using the
cases $(n,m)$ and $(1,m)$ (the case $(n,m+1)$ is completely
symmetric): \def\fig{isos-are-isos-boxempty-prf-induction}
\begin{align*}
\InputIfFileExists{./figures/\fig/\fig_00.tikz}{}{{\color{red}\colorbox{pink}{missing file : 00}}}
~\equiv~\InputIfFileExists{./figures/\fig/\fig_01.tikz}{}{{\color{red}\colorbox{pink}{missing file : 01}}}\\
~~\equiv~~\InputIfFileExists{./figures/\fig/\fig_02.tikz}{}{{\color{red}\colorbox{pink}{missing file : 02}}}
~~\equiv~~\InputIfFileExists{./figures/\fig/\fig_03.tikz}{}{{\color{red}\colorbox{pink}{missing file : 03}}}
\end{align*}
In a similar way, it is possible to show the following three identities:
\[\tikzfig{isos-are-isos-boxempty-2}\]
when all the wires $a_{ij}, b_i,c_j$ are mutually exclusive.

\[\tikzfig{isos-are-isos-tensor}\] when all the wires $a_{ij}$.

\[\tikzfig{isos-are-isos-tensor-2}\] when all the wires $a_{ij}$ are
mutually exclusive.
\begin{itemize}
\item~[$\iso_{\mathcal A}^{\inv}\circ\iso_{\mathcal A}$]: The result
is obvious in cases $\one$, $\zero$ and $\emptysquare$. For $A\oplus B$:
\def\fig{isos-are-isos-plus}
\begin{align*}
\InputIfFileExists{./figures/\fig/\fig_00.tikz}{}{{\color{red}\colorbox{pink}{missing file : 00}}}
\equiv\InputIfFileExists{./figures/\fig/\fig_01.tikz}{}{{\color{red}\colorbox{pink}{missing file : 01}}}
\equiv\InputIfFileExists{./figures/\fig/\fig_02.tikz}{}{{\color{red}\colorbox{pink}{missing file : 02}}}
\equiv\InputIfFileExists{./figures/\fig/\fig_03.tikz}{}{{\color{red}\colorbox{pink}{missing file : 03}}}
\end{align*}
The proof is similar for $\tensor$ and $\parallel$ using the previous identities.
\item~[$\iso_{\mathcal A}\circ\iso_{\mathcal A}^{\inv}$]: The result
is again obvious for $\one$, $\zero$ and $\emptysquare$. The general result is
easy to prove by induction using the above identities.  \qedhere
\end{itemize}
\end{proof}

\subsection{Universality}
We are now ready to show the universality of the language:

\begin{thm}[Universality]~\\
	\label{thm:universality}
	For every objects $\mathcal{A},\mathcal{B}$ of $\catD$, and for
	every linear operator

	\begin{align*}
		\Lambda &\in \FdM\left(
		\bigoplus_{\mathcal{E} \in \mathcal{A}^{\bullet}}
		\mathcal{H}_{\mathcal{E}},
		\bigoplus_{\mathcal{F} \in \mathcal{B}^{\bullet}}
		\mathcal{H}_{\mathcal{B}}
		\right)\\
		\Lambda &= \begin{pmatrix}
		\lambda_{1\,1} & \cdots & \lambda_{n\,1} & \lambda_{10} \\
		\vdots & & \vdots & \vdots \\
		\lambda_{1\,m} & \cdots & \lambda_{n\,m} & \lambda_{m0} \\
		\lambda_{01} & \cdots & \lambda_{0n} & \lambda_{00}
		\end{pmatrix}
	\end{align*}
	there exists a set of worlds
	$W$, and a morphism $\withWorld{f}{W} \in
	\ManyWorlds{\forall}(\mathcal{A},\mathcal{B})$ such that its
	worlds-agnostic semantics $\interp{f}$ is equal to $U$.
\end{thm}

\begin{proof}
We take $f = \iso_{\mathcal{B}}^{-1} \circ \lambda \circ
\iso_{\mathcal{A}}$ as defined above and show that $\interp{f} =
\Lambda$. As stated in Section~\ref{sec:normal}, there is one world for each
scalar in $\lambda$, so we write $a_{i\,j}$ the world associated to
$\lambda_{i\,j}$  and $W$ the set of all those worlds. Let us write
$\one^k_0=\bullet^{k}$ for $\bullet \parallel \dots \parallel \bullet$
(with $k$ elements) and $\one^k_i$ for $\bullet^k$ where the $i$-th
$\bullet$ has been replaced by $\one$ if $1\leq i\leq k$. Beware that
we consider $\one^k_0$ to be the \emph{last} element of the canonical
basis. We then have $\interp{\lambda}_{a_{i\,j}} :\one^n_i \to
\one^m_j : x \mapsto \lambda_{i\,j}  \cdot x$ where:
\[\tikzfig{NF-matrix-block-gen}\] Additionally, one can show by
induction that $\interp{\iso_{\mathcal{A}}}_{a_{i\,j}}$ is simply the
projection on the $i$-th element of the canonical basis, and
$\interp{\iso^{-1}_{\mathcal{B}}}_{a_{i\,j}}$ is simply the injection on
the $j$-th element of the canonical basis. Since we have $\interp{f}_a
= \interp{\iso^{-1}_{\mathcal{B}}}_{a} \circ \interp{\lambda}_a \circ
\interp{\iso_{\mathcal{A}}}_{a}$ for every $a \in W$, and $\interp{f}$
being the collection of all the $\interp{f}$, we obtain that
$\interp{f} = \Lambda$.
\end{proof}

\subsection{Uniqueness of the Normal Form}

A crucial feature of the normal form for the completeness is its
uniqueness:
\begin{prop}
The normal form is unique.
\end{prop}

\begin{proof}
Let $f$ and $g$ be two diagrams in normal form (with respectively
$\lambda$ and $\mu$ as inner block), such that $\interp{f}=\interp{g}$
(the naming of the worlds is taken to be the same in both diagrams,
and is the same as in the previous proof). By the definition of
$\interp{.}$, we have $\interp{f}_a=\interp{g}_a$ for every $a\in W$.
We hence have $\interp{\iso^{-1}_{\mathcal{B}}}_{a} \circ
\interp{\lambda}_a \circ \interp{\iso_{\mathcal{A}}}_{a} =
\interp{\iso^{-1}_{\mathcal{B}}}_{a} \circ \interp{\mu}_a \circ
\interp{\iso_{\mathcal{A}}}_{a}$.

Denoting $e_i^{\mathcal A}$ (resp.~$e_i^{\mathcal B}$) the $i$-th
element of the basis of $\mathcal A$ (resp.~$\mathcal B$), we have:
\begin{align*}
\lambda_{ij}
&= e_j^{\mathcal
B\dagger}\interp{\iso^{-1}_{\mathcal{B}}}_{a_{ij}} \circ
\interp{\lambda}_{a_{ij}} \circ
\interp{\iso_{\mathcal{A}}}_{a_{ij}}e_i^{\mathcal A}\\
&= e_j^{\mathcal B\dagger}\interp{\iso^{-1}_{\mathcal{B}}}_{a_{ij}} \circ
\interp{\mu}_{a_{ij}} \circ
\interp{\iso_{\mathcal{A}}}_{a_{ij}}e_i^{\mathcal A}
= \mu_{ij}
\end{align*}
Hence, all
coefficients in the scalars of $f$ and $g$ are the same. Since the
structure is otherwise the same for $f$ and $g$, they are the same
diagram.
\end{proof}

\subsection{Completeness}
\label{subseq:completeness}

We can now use this normal form to show that our equational theory is
complete for arbitrary morphisms. To do so, we need to show that all
the generators can be put in normal form, and then that any
composition of morphisms in normal form can be put in normal form.
To do so we will first derive a few lemmas:

\begin{cor}[of Lemma~\ref{lem:contraction-natural}]
	\phantomsection\label{lem:interversion-unitor-contraction}
	\def\fig{lemma-contraction-unitor-exchange}
	$\InputIfFileExists{./figures/\fig/\fig_00.tikz}{}{{\color{red}\colorbox{pink}{missing file : 00}}}
	\equiv\InputIfFileExists{./figures/\fig/\fig_05.tikz}{}{{\color{red}\colorbox{pink}{missing file : 05}}}$ when $s_1\cap s_4 = s_2\cap s_3 = \varnothing$
\end{cor}

\begin{cor}[of Lemma~\ref{lem:contraction-natural}]
	\label{lem:isos-distribute-over-c}
	Single-colored isos distribute over the contraction:
	\[\tikzfig{isos-distribute-over-c}\]
\end{cor}

\begin{cor}[of Lemma~\ref{lem:contraction-natural}]
	\label{lem:matrix-blocks-distribute-over-c}
	\[\tikzfig{NF-matrix-block-copy-lem}\]
\end{cor}

\begin{lem}
	\label{lem:scalars-distribute-over-isos}
	Scalars distribute over single-colored isos:
	\[\tikzfig{scalars-distribute-over-isos}\]
\end{lem}

\begin{proof}
\phantomsection\label{prf:scalars-distribute-over-isos}
The result is obvious for $\one$ and $\emptysquare$. For $A\oplus B$:
\def\fig{scalars-distribute-over-isos-prf-oplus}
\begin{align*}
\InputIfFileExists{./figures/\fig/\fig_00.tikz}{}{{\color{red}\colorbox{pink}{missing file : 00}}}
=\InputIfFileExists{./figures/\fig/\fig_01.tikz}{}{{\color{red}\colorbox{pink}{missing file : 01}}}
&\equiv\InputIfFileExists{./figures/\fig/\fig_02.tikz}{}{{\color{red}\colorbox{pink}{missing file : 02}}}
\equiv\InputIfFileExists{./figures/\fig/\fig_03.tikz}{}{{\color{red}\colorbox{pink}{missing file : 03}}}
=\InputIfFileExists{./figures/\fig/\fig_04.tikz}{}{{\color{red}\colorbox{pink}{missing file : 04}}}
\end{align*}
For $A\tensor B$:
\def\fig{scalars-distribute-over-isos-prf-otimes}
\begin{align*}
\InputIfFileExists{./figures/\fig/\fig_00.tikz}{}{{\color{red}\colorbox{pink}{missing file : 00}}}
&=\InputIfFileExists{./figures/\fig/\fig_01.tikz}{}{{\color{red}\colorbox{pink}{missing file : 01}}}
\equiv\InputIfFileExists{./figures/\fig/\fig_02.tikz}{}{{\color{red}\colorbox{pink}{missing file : 02}}}
\equiv\InputIfFileExists{./figures/\fig/\fig_03.tikz}{}{{\color{red}\colorbox{pink}{missing file : 03}}}
\equiv\InputIfFileExists{./figures/\fig/\fig_04.tikz}{}{{\color{red}\colorbox{pink}{missing file : 04}}}\\
&\equiv\InputIfFileExists{./figures/\fig/\fig_05.tikz}{}{{\color{red}\colorbox{pink}{missing file : 05}}}
=\InputIfFileExists{./figures/\fig/\fig_06.tikz}{}{{\color{red}\colorbox{pink}{missing file : 06}}}
\end{align*}
\end{proof}

\begin{cor}[of Lemma~\ref{lem:contraction-natural}]
	\label{lem:dangling-unitor-distribution}
	\def\fig{lemma-dangling-unitor-distribution}
	\begin{align*}
	\InputIfFileExists{./figures/\fig/\fig_00.tikz}{}{{\color{red}\colorbox{pink}{missing file : 00}}}
	\equiv\InputIfFileExists{./figures/\fig/\fig_03.tikz}{}{{\color{red}\colorbox{pink}{missing file : 03}}}
	\end{align*}
	with $s_0\cap s_1=\varnothing$.
\end{cor}

\begin{lem}
	\label{lem:sequential-composition-of-matrix-blocks}
	\def\fig{NF-matrix-block-gen-compo}
	\begin{align*}
	&\InputIfFileExists{./figures/\fig/\fig_00.tikz}{}{{\color{red}\colorbox{pink}{missing file : 00}}}
	\equiv\InputIfFileExists{./figures/\fig/\fig_05.tikz}{}{{\color{red}\colorbox{pink}{missing file : 05}}}
	\end{align*}
	with $\nu_{ij} = \sum_k\lambda_{ik}\mu_{kj}$.
\end{lem}

\begin{proof}
\phantomsection\label{prf:sequential-composition-of-matrix-blocks}
Suppose the worlds of $\lambda$ are the $\{a_{ij}\}_{ij}$ and that of
$\mu$ are the $\{b_{k\ell}\}_{k\ell}$. We count inputs/outputs
starting at $1$, hence $p\geq1$ in the following.

 The $p$-th output of $\lambda$ has world set $\{a_{ip}\}_{i}$ and the
 $p$-th input of $\mu$ has world set $\{b_{p\ell}\}_{\ell}$. When
 composing the two in sequence, in the first step, each singleton
 world $\{a_{ij}\}$ (resp.~$\{b_{k\ell}\}$) is mapped to
 $\{(a_{ij},b_{k\ell})\}_{k\ell}$
 (resp.~$\{(a_{ij},b_{k\ell})\}_{ij}$), and unions of singleton worlds
 to unions of the worlds each is mapped to.

In particular, the $p$-th output of $\lambda$ now has world set
$\{(a_{ip},b_{k\ell})\}_{ik\ell}$, and the $p$-th input of $\mu$ now
has world set $\{(a_{ij},b_{p\ell})\}_{ij\ell}$. After composition,
the two sets have to match. They become
$\{(a_{ip},b_{p\ell})\}_{i\ell}$, and all pairs of world that were
removed from these two sets are removed globally.

We do so for all $p\geq 1$, which means all worlds in:
\begin{align*}
\{(a_{ij},b_{k\ell})\mid j\geq1, j\neq
k\}\cup\{(a_{ij},b_{k\ell})\mid k\geq1, j\neq
k\}\\
=\{(a_{ij},b_{k\ell})\mid j\neq k\}
\end{align*}
are removed. Crucially, this
means that the world sets $\{a_{i0}\}_{i}$ and $\{b_{0\ell}\}_{\ell}$
are mapped to the same world set:
\begin{align*}
\{a_{i0}\}_{i}
\mapsto &~\{(a_{i0},b_{k\ell})\}_{ik\ell}\setminus\{(a_{ij},b_{k\ell})\mid j\neq k\}=\{(a_{i0},b_{0\ell})\}_{i\ell}\\
&
=\{(a_{ij},b_{0\ell})\}_{ij\ell}\setminus\{(a_{ij},b_{k\ell})\mid j\neq k\}
\mapsfrom\{b_{0\ell}\}_{\ell}
\end{align*}

Hence, the sequential composition of $\lambda$ and $\mu$ becomes:
\def\fig{NF-matrix-block-gen-alt-compo-aux}
\begin{align*}
\InputIfFileExists{./figures/\fig/\fig_00.tikz}{}{{\color{red}\colorbox{pink}{missing file : 00}}}
&\equiv\InputIfFileExists{./figures/\fig/\fig_01.tikz}{}{{\color{red}\colorbox{pink}{missing file : 01}}}
\equiv\InputIfFileExists{./figures/\fig/\fig_02.tikz}{}{{\color{red}\colorbox{pink}{missing file : 02}}}\\
&\equiv\InputIfFileExists{./figures/\fig/\fig_03.tikz}{}{{\color{red}\colorbox{pink}{missing file : 03}}}
\equiv\InputIfFileExists{./figures/\fig/\fig_04.tikz}{}{{\color{red}\colorbox{pink}{missing file : 04}}}
\end{align*}
with $\nu_{ij} = \sum_k\lambda_{ik}\mu_{kj}$.
\end{proof}

\begin{lem}
	\label{lem:matrix-block-through-iso-parallel}
	For any ``matrix block'' $\lambda$, there exists a ``matrix block'' $\nu$ such that:
	\def\fig{NF-matrix-block-through-iso-boxempty}
	\begin{align*}
	\InputIfFileExists{./figures/\fig/\fig_00.tikz}{}{{\color{red}\colorbox{pink}{missing file : 00}}}
	\equiv\InputIfFileExists{./figures/\fig/\fig_05.tikz}{}{{\color{red}\colorbox{pink}{missing file : 05}}}
	\end{align*}
\end{lem}

\begin{proof}
First, notice the following:
\def\fig{NF-tensor-lemma}
\begin{align*}
\label{eq:NF-tensor-lemma}
\InputIfFileExists{./figures/\fig/\fig_01.tikz}{}{{\color{red}\colorbox{pink}{missing file : 01}}}
\equiv\InputIfFileExists{./figures/\fig/\fig_02.tikz}{}{{\color{red}\colorbox{pink}{missing file : 02}}}
\equiv\InputIfFileExists{./figures/\fig/\fig_03.tikz}{}{{\color{red}\colorbox{pink}{missing file : 03}}}
\equiv\InputIfFileExists{./figures/\fig/\fig_04.tikz}{}{{\color{red}\colorbox{pink}{missing file : 04}}}\addtocounter{equation}{1}\tag{\theequation}
\end{align*}
We may then prove the lemma:
\def\fig{NF-matrix-block-through-iso-boxempty-gen}
\begin{align*}
\InputIfFileExists{./figures/\fig/\fig_00.tikz}{}{{\color{red}\colorbox{pink}{missing file : 00}}}\hspace*{-4em}&\hspace*{4em}
=\InputIfFileExists{./figures/\fig/\fig_01.tikz}{}{{\color{red}\colorbox{pink}{missing file : 01}}}
\equiv\InputIfFileExists{./figures/\fig/\fig_02.tikz}{}{{\color{red}\colorbox{pink}{missing file : 02}}}\\
&\underset{\ref{lem:matrix-blocks-distribute-over-c}}{\equiv}\InputIfFileExists{./figures/\fig/\fig_03.tikz}{}{{\color{red}\colorbox{pink}{missing file : 03}}}
\underset{\ref{eq:NF-tensor-lemma}}\equiv\InputIfFileExists{./figures/\fig/\fig_04.tikz}{}{{\color{red}\colorbox{pink}{missing file : 04}}}\\
&\equiv\InputIfFileExists{./figures/\fig/\fig_05.tikz}{}{{\color{red}\colorbox{pink}{missing file : 05}}}
\equiv\InputIfFileExists{./figures/\fig/\fig_06.tikz}{}{{\color{red}\colorbox{pink}{missing file : 06}}}\\
&=\InputIfFileExists{./figures/\fig/\fig_07.tikz}{}{{\color{red}\colorbox{pink}{missing file : 07}}} \tag*{\qed}
\end{align*}
\renewcommand{\qed}{}
\end{proof}

We can now move on to show that generators can be put in normal form:

\begin{prop}
	\label{prop:NF-generators}
	The generators can be put in normal form.
\end{prop}

\begin{proof}
	\phantomsection\label{prf:NF-generators}
First for the tensor:
	\[\tikzfig{NF-tensor-simp-gen}\]
Then, for the plus:
	\[\tikzfig{NF-plus-simp-gen}\]
	\def\fig{NF-swap-boxempty}
The swap:
\begin{align*}
\InputIfFileExists{./figures/\fig/\fig_00.tikz}{}{{\color{red}\colorbox{pink}{missing file : 00}}}
&~\equiv~\InputIfFileExists{./figures/\fig/\fig_01.tikz}{}{{\color{red}\colorbox{pink}{missing file : 01}}}
~\equiv~\InputIfFileExists{./figures/\fig/\fig_02.tikz}{}{{\color{red}\colorbox{pink}{missing file : 02}}}
~\equiv~\InputIfFileExists{./figures/\fig/\fig_03.tikz}{}{{\color{red}\colorbox{pink}{missing file : 03}}}
~\equiv~\InputIfFileExists{./figures/\fig/\fig_04.tikz}{}{{\color{red}\colorbox{pink}{missing file : 04}}}
\end{align*}
	with $\sigma$ a simple permutation of wires.
	For the case of the contraction, the unary case is dealt with thanks to Lemma~\ref{lem:contraction-zero}. The binary case is dealt with:
	\[\tikzfig{NF-contraction-gen}\]
	For the general, $n$-ary case of the contraction, it suffices to decompose any contraction into binary contractions, and use the above equality.
	\[\tikzfig{NF-scalar}\]
	This normal form encompasses that of the identity, by simply taking $s = 1$. The unit is simply obtained as:
	\[\tikzfig{NF-unit}\]
	Finally, the cap is obtained as follows:
	\def\fig{NF-cap}
\begin{align*}
\InputIfFileExists{./figures/\fig/\fig_00.tikz}{}{{\color{red}\colorbox{pink}{missing file : 00}}}
&~\equiv~\InputIfFileExists{./figures/\fig/\fig_01.tikz}{}{{\color{red}\colorbox{pink}{missing file : 01}}}
~\equiv~\InputIfFileExists{./figures/\fig/\fig_02.tikz}{}{{\color{red}\colorbox{pink}{missing file : 02}}}
~\equiv~\InputIfFileExists{./figures/\fig/\fig_03.tikz}{}{{\color{red}\colorbox{pink}{missing file : 03}}}
~\equiv~\InputIfFileExists{./figures/\fig/\fig_04.tikz}{}{{\color{red}\colorbox{pink}{missing file : 04}}}
\end{align*}
	The upside-down versions of the generators
	are provided in exactly the same way (but upside-down).
 \end{proof}

\begin{prop}
	\label{prop:NF-compo}
	Compositions of diagrams in normal form can be put in normal form.
\end{prop}

It is then possible to show that compositions of diagrams in normal
form can be put in normal form:
\begin{proof}
	\phantomsection\label{prf:NF-compo}
	In the case of sequential composition:
	\def\fig{NF-compo-circ-gen}
	\begin{align*}
	\InputIfFileExists{./figures/\fig/\fig_00.tikz}{}{{\color{red}\colorbox{pink}{missing file : 00}}}
	\underset{\ref{lem:isos-are-isos}}\equiv\InputIfFileExists{./figures/\fig/\fig_01.tikz}{}{{\color{red}\colorbox{pink}{missing file : 01}}}
	\underset{\ref{lem:sequential-composition-of-matrix-blocks}}\equiv\InputIfFileExists{./figures/\fig/\fig_02.tikz}{}{{\color{red}\colorbox{pink}{missing file : 02}}}
	\end{align*}
	In the case of parallel composition:
	\def\fig{NF-compo-boxempty}
	\begin{align*}
	&\InputIfFileExists{./figures/\fig/\fig_00.tikz}{}{{\color{red}\colorbox{pink}{missing file : 00}}}
	\underset{\ref{lem:isos-are-isos}}\equiv\InputIfFileExists{./figures/\fig/\fig_01.tikz}{}{{\color{red}\colorbox{pink}{missing file : 01}}}
	\underset{\ref{lem:isos-are-isos}}\equiv\InputIfFileExists{./figures/\fig/\fig_02.tikz}{}{{\color{red}\colorbox{pink}{missing file : 02}}}
	\underset{\ref{lem:matrix-block-through-iso-parallel}}\equiv\InputIfFileExists{./figures/\fig/\fig_03.tikz}{}{{\color{red}\colorbox{pink}{missing file : 03}}}\\
	&\underset{\ref{lem:matrix-block-through-iso-parallel}}\equiv\InputIfFileExists{./figures/\fig/\fig_04.tikz}{}{{\color{red}\colorbox{pink}{missing file : 04}}}
	\underset{\substack{\ref{lem:isos-are-isos}\\\ref{lem:sequential-composition-of-matrix-blocks}}}\equiv\InputIfFileExists{./figures/\fig/\fig_05.tikz}{}{{\color{red}\colorbox{pink}{missing file : 05}}}
	\underset{\ref{lem:quasi-associativity-of-parallel}}\equiv\InputIfFileExists{./figures/\fig/\fig_06.tikz}{}{{\color{red}\colorbox{pink}{missing file : 06}}}
	\underset{\ref{lem:sequential-composition-of-matrix-blocks}}\equiv\InputIfFileExists{./figures/\fig/\fig_07.tikz}{}{{\color{red}\colorbox{pink}{missing file : 07}}}
	\end{align*}
	where $\overline{\mathcal A\parallel \mathcal C}$ represents the
	canonical choice of composition with $\parallel$ (and similarly
	for $\overline{\mathcal B\parallel \mathcal D}$).
\end{proof}

We are now ready to show the completeness of the Many-Worlds Calculus:

\begin{thm}[Completeness]\label{thm:completeness} For every $\withWorld{f}{W}
	: {\mathcal{A}} \to {\mathcal{B}}$ and $\withWorld{g}{V}: {\mathcal{A}} \to
	{\mathcal{B}}$, $\interp{\withWorld{f}{W}} = \interp{\withWorld{g}{V}} \iff \withWorld{f}{W} \equiv
	\withWorld{g}{V}$.
\end{thm}
\begin{proof}
	The right-to-left direction of the equivalence can be directly
        checked by verifying that all the axioms preserve the
        semantics.

	Let $f_1$ and $f_2$ be two morphisms such that
        $\interp{f_1}=\interp{f_2}$. Both morphisms can be put in
        normal form, resp.~$f_1^{NF}$ and $f_2^{NF}$, with $f_i\equiv
        f_i^{NF}$ and thus $\interp{f_i^{NF}}=\interp{f_i}$. By
        uniqueness of the normal form, and since
        $\interp{f_1^{NF}}=\interp{f_2^{NF}}$, we get $f_1^{NF}\equiv
        f_2^{NF}$, which ends the proof that $f_1\equiv f_2$.
\end{proof}

\section{Comparison with Other Graphical Languages}

The distinctive feature of the Many-Worlds Calculus is that it
graphically puts the tensor and the biproduct on an equal footing. By
comparison, other graphical language for quantum computing are
inherently centered around either one of them. The
ZX-calculus~\cite{zxorigin} and cousin languages ZW- and
ZH-Calculi~\cite{Backens2019ZH,Hadzihasanovic2015axiomatisation}, as
well as Duncan's Tensor-Sum Logic~\cite{duncan2009generalized}, use
the tensor product as the default monoid, while more recent language
-- particularly for linear optics~\cite{pbs,path,lov} -- use the
biproduct. We have a closer look at each of them in the following, and
show how -- at least part of -- each language can be encoded naturally
in the Many-Worlds Calculus. Most of them comes equipped with an
equational theory. By completeness of our language
(Theorem~\ref{thm:completeness}), all the equations expressible in
the fragments we consider can be derived in our framework.

\subsection{ZX-Calculus}

The first difference is the restrictions of the ZX-calculus to
computations between qubits, in other words linear map from
$\mathbb{C}^{2^n} \mapsto \mathbb{C}^{2^m}$, while our language can
encode any linear map from $\mathbb{C}^{n} \mapsto \mathbb{C}^{m}$.
The Tensor generator allowing the decomposition of $\mathbb{C}^{2^n}$ into
instances of $\mathbb{C}^2$ was already present in the \emph{scalable}
extension of the ZX-calculus~\cite{carette2019SZX}, but the main
difference comes from the Plus (and the Contraction).

Additionally, every ZX-diagram can be encoded in our graphical
language. The identity, swap, cup and cap of the ZX calculus are
encoded by the similar generators over the type $\one \oplus \one$,
the Hadamard gate is encoded as in Figure~\ref{fig:hadamard}, and the green
spider is encoded as shown below. An encoding for the red
spider can then be deduced from those. Diagrams are composed
together with the world-agnostic composition of
$\ManyWorlds{\forall}$.

\[\tikzfig{trad-zx-gn} \]

\subsection{Tensor-Sum Logic}

The core difference between the Tensor-Sum
Logic~\cite{duncan2009generalized} work and ours is the presence of
the contraction in our graphical language. They instead rely on an
enrichment of their category by a sum, which they represent
graphically with boxes. We show below how the morphism $f+g$
would be encoded in both their and our language. More generally, their
boxes correspond to uses of our contraction generator in a
``well-bracketed'' way. Another point of difference is their approach to
quantum computation, as we do not assign the same semantics to those
superpositions of morphisms. In their approach, the superposition is a
classical construction and corresponds to the measurement and the
classical control flow, while in our approach the superposition is a
quantum construction and corresponds to the quantum control.

\[\tikzfig{box-contraction}\]

\subsection{PBS-Calculus}
The PBS-Calculus~\cite{pbs} allows one to represent coherent quantum
control by the use of \emph{polarizing beam splitters} (pbs): whenever
a qubit enters a pbs node, depending on the polarity of the qubit it will
either go through or be reflected. By making implicit the target
system, controlled by the optical system represented by the diagram,
the PBS-Calculus allows one to encode the Quantum Switch (depicted on
the left). The pbs generator is related to the $\oplus$ of the
Many-Worlds.

\[\tikzfig{pbs} \qquad \qquad \tikzfig{pbs-to-mwc}\]

The first main difference with our language is that, since the
generators of the PBS-Calculus represent physical components, any
PBS-diagram is by construction physical, while our approch is more
atomic and decomposes physical components into abstract smaller ones.
The second main difference lie in the trace: while they can allow a
particle to pass through a wire at most twice, in our system, each
wire can be used at most once: more formally, their trace is based on
the coproduct while ours is on the tensor product. \emph{If} we are
assured that each wire can only be used once during the computation,
any PBS-diagram can be translated to the Many-Worlds calculus
directly, with the transformation on the right, where we distinguish
the control system (the part of the diagram connected to {\tikz
\node[style=plus] {};}\,s) from the target system (connected to {\tikz
\node[style=contraction] {};}\,s) which is implicit in the
PBS-Calculus. Finally, as we mentionned in
Section~\ref{sec:intro}, whereas the PBS-calculus uses black boxes,
there is no need for them in the Many-Worlds as they can directly be
encoded in our system. Also, the distinction between the control and
controlled system is no longer present, as they can change during the
computation.

\subsection{LOv-Calculus}

In the PBS-Calculus, the qubit in the control system (the one explicitly
represented) cannot be put in arbitrary superpositions of $\ket0$ and $\ket1$
\emph{during} the computation. To allow this feature, we may add some linear
optical components to the language's generators, and end up with the
LOv-Calculus~\cite{lov}. In this language, there is no trace and there is a
unique photon traveling the circuit, which relieves us of
the previous constraint. There is also no need for an implicit target system
anymore. All wires at the interface between the generators are of type
$\one\oplus\one$, and parallel wires have disjoint sets of worlds. Each
generator can then be interpreted as follows:
\[
\tikzfig{lov-to-mwc-wave-plates}
\qquad\quad
\tikzfig{lov-to-mwc-BS}\]
\[\tikzfig{lov-to-mwc-PBS}
\qquad\quad
\tikzfig{lov-to-mwc-phase-shifter}
\qquad\quad
\tikzfig{lov-to-mwc-vacuum-source}\]

\subsection{Path-Calculus}

The Path-Calculus is another recent graphical language for linear optical
circuits~\cite{path}. Its generators correspond directly to a subset of the
Many-Worlds' with \tikzfig{path-to-mwc-unary-contraction},
\tikzfig{path-to-mwc-binary-contraction} and \tikzfig{path-to-mwc-scalar};
where each wire has type $\one$ and where parallel wires are on disjoint sets
of worlds. This language is then used as the core for a more
expressive language called QPath, which this time cannot be directly encoded
in our language, except when restricting the set of generators (specifically
to $n=1$), in which case \tikzfig{path-to-mwc-ket1}.

\subsection{Tapes diagrams}

In \cite{Bonchi2022deconstructing}, \emph{tape diagrams} are
introduced to display both tensor product and biproduct, and a sound
and complete equational theory is presented. There, systems that are
in tensors are bundled in ``tapes'', and these tapes can then be put
side-by-side, representing the coproduct of the two tensors. This
framework turns out to be less verbose, since no world annotation is
needed, and the distinction between tensor and coproduct is completely
handled by the tapes. However, it forces the types to be fully
distributed, at any point, which may lead to an exponential blowup in
the number of tapes required. For example, the identity on
$(A\oplus B)\otimes C$ has to be distributed and is therefore a
diagram on $(A\otimes C) \oplus (A\otimes B)$, while in the
Many-Worlds it consists of a simple wire of the non-distributed type:
\[\scalebox{0.7}{\tikzfig{tapes}} \tikzfig{id-tapes-mw}\]

\subsection{Multiplicative Additive Proof Nets}

The syntax of the Many-Worlds is very close to the one of
multiplicative additive proof nets, in fact, the connexion between
proof nets and quantum computation has already been studied in the
multiplicative setting~\cite{duncan2004believe,duncan2006types} and
also in the Tensor-Sum Logic (that we mentioned
above)~\cite{duncan2009generalized}. The Many-Worlds Calculus acts as
a model of multiplicative additive linear logic~\cite{linearlogic},
where both additive connectives are collapsed to the $\oplus$ and both
multiplicative connectives are collapsed to the $\tensor$. Meanwhile,
the multiplicative (resp. additive) units are collapsed to the type
\one (resp. \zero). While the notion of worlds is very similar to the
notion of weights~\cite{proofnets, monomialPN}, the correspondence is
not exact, and the notion of validity criterion of proof nets such as
the one present in~\cite{hughes2003proof} cannot be applied to the
Many-Worlds Calculus (as everything is self-dual).

\subsection{Routed Circuits}
Routed circuits \cite{vanrietvelde2021routed} are a generalization of usual quantum circuits,
designed to accommodate indefinite causal orders. This setting allows for
``feedback loops'' (partial traces), and uses \emph{sectorial constraints}
on the different Hilbert spaces to ensure these do not break unitarity.
The quantum switch is the prime example of a non-causally ordered process
that can be expressed as a routed circuit. We have already shown how it can
also be represented as a Many-Worlds diagram. More generally,
we conjecture that all processes expressed as routed circuits can be turned
into Many-Worlds diagrams, in a way that preserves the semantics, by using
the worlds system to represent the sectorial constraints. While the precise
connection is left as future work, we illustrate it through an additional
example of a non-causally ordered process, namely the Lugano process
\cite{Baumeler2016space,vanrietvelde2023consistent}.
The Lugano process is a tri-partite process whose purpose is to elect a leader among the three parties. Each party can only vote for either of the two other parties, and the party that gets the majority is elected. What makes this process non-causally ordered, is the fact that each party knows whether they have won or not, \emph{before} they even cast their ballot. Literature shows that, surprisingly, this does not create a grandfather-like paradox, and that this technically classical process can be made quantum.
The Lugano process is described as a routed circuit as follows:
\def\fig{Lugano-2}
\begin{align*}
\operatorname{Lugano}(A,B,C)~:=~\InputIfFileExists{./figures/\fig/\fig_00.tikz}{}{{\color{red}\colorbox{pink}{missing file : 00}}}
\end{align*}
where the boxes $A$, $B$ and $C$ are the parameters of the process, all the wires except the first output are of type $\one\oplus\one$, and the sectorial constraints are implicitly contained in the other components that are given a Many-Worlds representation in Figure~\ref{fig:Lugano-components} (all wires in the above diagram have worlds set $\{a,a',b,b',c,c'\}$ in accordance with the legend in this figure).

\begin{figure}[!htb]
\begin{tabular}{|c|}
\hline
Legend\\[0.5em]
\tikzfig{Lugano-legend}\\
\hline
\end{tabular}
\begin{tabular}{c}
\tikzfig{Lugano-V}\qquad\tikzfig{Lugano-W}\\
\tikzfig{Lugano-CX}\\~
\end{tabular}
\tikzfig{Lugano-F}
\caption{Components of the Lugano process. $V_i$, $W_i$ and CX$_i$ for $i\in\{b,c\}$ are defined similarly by adequately permuting the worlds $a$, $a'$, $b$, $b'$, $c$ and $c'$. Box $X$ is the usual Not gate, as represented e.g.~in Figure~\ref{fig:ex_CNOT_decomposed}. We keep it as a box here simply to avoid having to use yet another 6 colors (for worlds sets $\{a\}$, $\{a'\}$, $\{b\}$, $\{b'\}$, $\{c\}$ and $\{c'\}$).}
\label{fig:Lugano-components}
\end{figure}

Using the equational theory, one can simplify the diagram, especially by making use of wires of type $\one$, and using the same color code as in Figure~\ref{fig:Lugano-components} to represent worlds sets:
\def\fig{Lugano-2}
\begin{align*}
\operatorname{Lugano}(A,B,C)
~\withWorld{\equiv}{W}\InputIfFileExists{./figures/\fig/\fig_01.tikz}{}{{\color{red}\colorbox{pink}{missing file : 01}}}
\hspace*{-1em}
\withWorld{\equiv}{W}\InputIfFileExists{./figures/\fig/\fig_02.tikz}{}{{\color{red}\colorbox{pink}{missing file : 02}}}
\end{align*}
The intuition behind the behavior of the process can be recovered once we make sense of the different world sets involved. In particular, $\{c,c',a\}$ can be understood as ``A votes for C'', and similarly for all the world sets used in boxes $V_i$. Then, worlds set $\{a,a'\}$ can be understood as ``A wins the vote'', which makes sense as $\{a,a'\} = \{a,a',b\}\cap\{a,a',c'\}$ i.e.~``C votes for A \emph{and} B votes for A''. The leftmost output then is a qutrit whose value indicates which of the three parties won the election.

\section{Conclusion}
We introduced a new sound and complete graphical language based on
compact categories with biproducts, along with an equational theory and
a worlds system, helping us to build a denotational semantics of our
language.

This language is a first step towards the unification of languages
based on the tensor $\otimes$ and those based on the biproduct
$\oplus$. This allows us to reason about both systems in parallel, and
superposition of executions, as shown by the encoding of the Quantum
Switch in Example~\ref{exa:qs} and the translation from term of the
language in Section~\ref{sec:isos-language}.

Following this translation, a natural development of the Many-Worlds
calculus consists in accommodating function abstraction and recursion
in the language. The question of a complete equational theory for the
language on mixed states (e.g.~via the discard construction
\cite{carette2019completeness}) is also left open.

Finally, while our language allows for quantum control and can faithfully represent the Quantum Switch, it does not
entirely capture another language that aims at formalizing quantum
control, namely the PBS-Calculus~\cite{pbs}. How and in which context
could we capture the PBS-Calculus is left for future work.

\section*{Acknowledgement}

This work has been partially funded by the French National Research
Agency (ANR) by the project RECIPROG ANR-21-CE48-0019, PPS
ANR-19-CE48-0014, TaQC ANR-22-CE47-0012 and within the framework of
``Plan France 2030'', under the research projects EPIQ
ANR-22-PETQ-0007, OQULUS ANR-22-PETQ-0013 and HQI-R\&D
ANR-22-PNCQ-0002. This work has also been partially funded by MIUR FARE project CAFFEINE, ``Compositional
and Effectful Program Distances'', R20LW7EJ7L.

\bibliographystyle{alphaurl}
\bibliography{bibli.bib}

\end{document}

%% file: macros.tex
\newcommand{\one}{\ensuremath{\mathds{1}}}
\DeclareMathAlphabet{\mymathbb}{U}{BOONDOX-ds}{m}{n}
\newcommand{\zero}{\ensuremath{\mymathbb{0}}}

\newcommand{\inl}[1]{\ensuremath{\mathtt{inj}_l{\;#1}}}
\newcommand{\inr}[1]{\ensuremath{\mathtt{inj}_r{\;#1}}}
\newcommand{\pv}[2]{\ensuremath{\langle #1,#2 \rangle}}

\newcommand{\set}[1]{\ensuremath{\{#1\}}}

\newsavebox\MBox

\usetikzlibrary{decorations.markings,arrows,decorations.pathmorphing, decorations.pathreplacing}
\usetikzlibrary{shapes.misc,matrix,backgrounds,folding,positioning}
\usetikzlibrary{shapes.geometric}
\usetikzlibrary{shapes.multipart}
\pgfdeclarelayer{edgelayer}
\pgfdeclarelayer{nodelayer}
\pgfsetlayers{background,edgelayer,nodelayer,main}

\tikzset{every path/.style={draw=black!80, line width=0.6pt}}
\tikzstyle{every picture}=[baseline=-0.25em]
\tikzstyle{none}=[inner sep=0mm]
\tikzstyle{box}=[fill=white, draw=black, shape=rectangle]
\tikzstyle{zxnode}=[shape=circle, minimum width=.25cm, inner sep=0.5pt, font=\footnotesize, draw=black,thick]
\tikzstyle{gn}=[zxnode ,fill=green, draw=green!10!black]
\tikzstyle{rn}=[zxnode ,fill=red, draw=red!10!black]
\tikzstyle{H box}=[rectangle,fill=yellow, draw=yellow!10!black,thick,xscale=1,yscale=1,font=\footnotesize,inner sep=1.2pt,minimum width=0.15cm,minimum height=0.15cm]
\tikzstyle{ug}=[regular polygon, regular polygon sides=3, fill=red,draw=black,inner sep = 0pt,minimum width=0.8em]
\tikzstyle{black dot}=[inner sep=0.4mm,minimum width=0pt,minimum height=0pt,fill=black,draw=black,shape=circle]
\tikzstyle{dot}=[black dot]
\tikzstyle{white dot}=[dot,fill=white, inner sep=-1pt, font=\footnotesize]
\tikzstyle{arrow}=[decoration={markings,mark=at position 1 with
    {\arrow[scale=1.2,>=stealth]{>}}},postaction={decorate}]

\tikzstyle{glabel}=[rounded corners=0.2em,fill=green!30,inner sep=0.1em,font=\scriptsize, anchor=west, xshift=-0.3em, yshift=0,opacity=1]
\tikzstyle{rlabel}=[rounded corners=0.2em,fill=red!30,inner sep=0.1em,font=\scriptsize, anchor=west, xshift=-0.3em, yshift=0,opacity=1]
\tikzstyle{box}=[rectangle, draw=black, fill=white, inner sep=1pt, font=\scriptsize]
\tikzstyle{box-no-outline}=[rectangle, draw=white, fill=white, inner sep=2pt]
\tikzstyle{circle-no-outline}=[circle, draw=white, fill=white, inner sep=0pt]
\tikzstyle{squigglearrow}=[->, line join=round, decorate, decoration={zigzag, segment length=4, amplitude=0.8, post=lineto, post length=2pt}]
\tikzstyle{divide}=[regular polygon, regular polygon sides=3, draw=black, fill=gray!50, inner sep=1.6pt, rounded corners=0.8mm]
\tikzstyle{very thick}=[-, line width=1pt]
\tikzstyle{boxedge}=[draw=gray!50]
\tikzstyle{pbs}=[diamond, draw=black, inner sep=-0.5pt, fill=white]
\tikzstyle{ribbon}=[thick, rounded corners=0.4pt,fill={rgb,255: red,157; green,246; blue,255}, fill opacity=0.7]
\tikzset{tensor/.style={inner sep=2.5pt, draw=.!70!white, fill=.!70!white, circle, path picture={
			\draw[white]
			(path picture bounding box.south east) -- (path picture bounding box.north west) (path picture bounding box.south west) -- (path picture bounding box.north east);
}}}
\tikzset{plus/.style={inner sep=2.5pt, draw, circle, path picture={
  \draw[black]
(path picture bounding box.east) -- (path picture bounding box.west) (path picture bounding box.south) -- (path picture bounding box.north);
}}}
\tikzstyle{contraction}=[circle,draw,font={\scriptsize \color{gray} c}, inner sep= 1pt]
\tikzstyle{small-contraction}=[circle,draw,font={\tiny  \color{gray} c}, inner sep= 0.2pt]
\tikzstyle{vacuum}=[rounded rectangle, draw, fill=gray!50, rounded rectangle west arc=none, rotate=180]
\tikzstyle{compact dash}=[dash pattern={on 2pt off 1pt}]

\pgfdeclarelayer{edgelayer}
\pgfdeclarelayer{nodelayer}
\pgfsetlayers{background,edgelayer,nodelayer,main}

\tikzstyle{every loop}=[]

\usetikzlibrary{circuits.ee.IEC}

\colorlet{tapeBg}{red!30}
\colorlet{tapeBorder}{red!60}
\tikzstyle{tapeFill} = [fill=tapeBg]
\tikzstyle{tapeNoFill} = [draw=tapeBorder, line width=0.5pt]
\tikzstyle{tape} = [fill=tapeBg, draw=tapeBorder, line width=0.5pt]

\newcommand{\tikzfig}[1]{
\InputIfFileExists{./figures/#1.tikz}{}{{\color{red}\colorbox{pink}{missing file : #1}}}
}

\def\fig{}

\iftoggle{extern}{
	\usetikzlibrary{external}
	\tikzexternalize[prefix=tikzfigs/]

	\renewcommand{\tikzfig}[1]{
		\tikzsetnextfilename{#1}
		
\InputIfFileExists{./figures/#1.tikz}{}{{\color{red}\colorbox{pink}{missing file : #1}}}
}

}{
	
}

\newcommand{\ket}[1]{\ensuremath{\left|  #1 \right\rangle}}

\newcommand{\cat}[1]{\ensuremath{\mathbf{#1}}}
\newcommand{\catD}{\ensuremath{\cat{C_D}}}
\newcommand{\interp}[1] {\left\llbracket #1 \right\rrbracket}

\usetikzlibrary{cd}

\let\boxempty\relax
\DeclareMathOperator*{\boxempty}{\parallel}

\newcommand{\ManyWorlds}[1]{\textbf{\textup{MW}}_{#1}}

\newcommand{\bfup}[1]{\textup{\textbf{#1}}}
\newcommand{\id}{\textup{\textbf{id}}}
\newcommand{\D}{\mathcal{D}}

\newcommand{\FdM}{\mathbf{FdS}_R}
\newcommand{\withWorld}[2]{{#1}_{\scaleto{#2}{4pt}}}

\newcommand{\emptysquare}{\ensuremath{\varnothing}}

\DeclareMathOperator{\tensor}{
\text{\scalerel*{%
		\iftoggle{extern}{\tikzsetnextfilename{tensor}}{}
		\begin{tikzpicture}
		\node[inner sep=2pt, draw=.!70!white, circle, fill=.!70!white, path picture={
			\draw[white]
			(path picture bounding box.south east) -- (path picture bounding box.north west) (path picture bounding box.south west) -- (path picture bounding box.north east);
		}] (c) at (0,0){};
		\end{tikzpicture}}{\otimes}}}

\DeclareMathOperator{\contraction}{
\scalerel*{%
		\iftoggle{extern}{\tikzsetnextfilename{contraction}}{}
		\begin{tikzpicture}
		\node[inner sep=2pt, draw=., circle, font={c}] (c) at (0,0){};
		\end{tikzpicture}}{\oplus}}

\newcommand{\iso}{\operatorname{iso}}
\newcommand{\inv}{\text-\scalebox{0.7}{$1$}}

\newcommand{\enab}[1]{#1^\bullet}
\newcommand{\enabW}[2]{{#1}\cap #2}

\renewcommand{\aa}{\textcolor{black!50!green}{a}}
\newcommand{\bb}{\textcolor{red}{b}}
\newcommand{\xx}{\textcolor{violet}{x}}
\newcommand{\cblue}{\textcolor{blue}{c}}
\newcommand{\sstar}{\textcolor{gray}{\star}}



\newcommand{\alt}{~\mid~}

\newcommand{\isot}{\leftrightarrow}

\newcommand{\letpv}[3]{{\tt let}\,{\langle #1 \rangle}={#2}~{\tt in}~{#3}}
\newcommand{\letv}[3]{{\tt let}\,{#1}={#2}~{\tt in}~{#3}}
\newcommand{\isobasique}{\ensuremath{\{v_1~\isot~e_1 \alt \dots \alt v_n~\isot~e_n\}}}
\newcommand{\tc}{\mathtt{t}\!\mathtt{t}}
\newcommand{\fc}{\mathtt{f}\!\mathtt{f}}
\newcommand{\isoterm}{\omega}

\newcommand{\opn}[1]{\ensuremath{\operatorname{#1}}}
\DeclarePairedDelimiter\pparenthesis{\llparenthesis}{\rrparenthesis}

%% file: figures/Lugano-2/Lugano-2_00.tikz
\begin{tikzpicture}
	\begin{pgfonlayer}{nodelayer}
		\node [style=none] (0)  at (-1.75, 0.5) {};
		\node [style=none] (1)  at (-1.0, 0.5) {};
		\node [style=none] (2)  at (-1.75, 1.0) {};
		\node [style=none] (3)  at (-1.0, 1.0) {};
		\node [style=none] (4)  at (-1.375, 0.75) {$V_a$};
		\node [style=none] (5)  at (-1.125, 0.5) {};
		\node [style=none] (6)  at (-1.625, 0.5) {};
		\node [style=none] (7)  at (-1.375, 1.0) {};
		\node [style=none] (8)  at (-2.0, 2.0) {};
		\node [style=none] (9)  at (-1.25, 2.0) {};
		\node [style=none] (10)  at (-2.0, 2.5) {};
		\node [style=none] (11)  at (-1.25, 2.5) {};
		\node [style=none] (12)  at (-1.625, 2.25) {CX$_a$};
		\node [style=none] (13)  at (-1.375, 2.0) {};
		\node [style=none] (14)  at (-1.875, 2.0) {};
		\node [style=none] (15)  at (-1.375, 2.5) {};
		\node [style=none] (16)  at (-1.875, 2.5) {};
		\node [style=none] (17)  at (-1.75, -1.25) {};
		\node [style=none] (18)  at (-1.0, -1.25) {};
		\node [style=none] (19)  at (-1.75, -0.75) {};
		\node [style=none] (20)  at (-1.0, -0.75) {};
		\node [style=none] (21)  at (-1.375, -1.0) {$W_a$};
		\node [style=none] (22)  at (-1.125, -1.25) {};
		\node [style=none] (23)  at (-1.625, -1.25) {};
		\node [style=none] (24)  at (-1.125, -0.75) {};
		\node [style=none] (25)  at (-1.625, -0.75) {};
		\node [style=none] (26)  at (-2.5, -2.75) {};
		\node [style=none] (27)  at (-2.5, -2.25) {};
		\node [style=none] (28)  at (2.5, -2.75) {};
		\node [style=none] (29)  at (2.5, -2.25) {};
		\node [style=none] (30)  at (0.0, -2.5) {$F$};
		\node [style=none] (31)  at (-0.25, -2.75) {};
		\node [style=none] (32)  at (-0.25, -3.0) {};
		\node [style=none] (33)  at (-2.25, -2.25) {};
		\node [style=none] (34)  at (0.25, -2.25) {};
		\node [style=none] (35)  at (-2.5, 2.5) {};
		\node [style=none] (36)  at (-2.5, -1.25) {};
		\node [style=none] (37)  at (-1.875, -1.25) {};
		\node [style=none] (38)  at (-1.375, 3.0) {};
		\node [style=none] (39)  at (-1.625, 1.75) {};
		\node [style=none] (40)  at (-1.625, 1.25) {};
		\node [style=none] (41)  at (-1.125, 1.25) {};
		\node [style=none] (42)  at (-1.125, 1.75) {};
		\node [style=none] (43)  at (-1.375, 1.25) {};
		\node [style=none] (44)  at (-1.375, 1.75) {};
		\node [style=none] (45)  at (-1.375, 1.5) {$A$};
		\node [style=none, font={\scriptsize}, text=black] (46)  at (-0.25, -3.25) {$\one\oplus\one$};
		\node [style=none] (47)  at (0.0, 0.5) {};
		\node [style=none] (48)  at (0.75, 0.5) {};
		\node [style=none] (49)  at (0.0, 1.0) {};
		\node [style=none] (50)  at (0.75, 1.0) {};
		\node [style=none] (51)  at (0.375, 0.75) {$V_b$};
		\node [style=none] (52)  at (0.625, 0.5) {};
		\node [style=none] (53)  at (0.125, 0.5) {};
		\node [style=none] (54)  at (0.375, 1.0) {};
		\node [style=none] (55)  at (-0.25, 2.0) {};
		\node [style=none] (56)  at (0.5, 2.0) {};
		\node [style=none] (57)  at (-0.25, 2.5) {};
		\node [style=none] (58)  at (0.5, 2.5) {};
		\node [style=none] (59)  at (0.125, 2.25) {CX$_b$};
		\node [style=none] (60)  at (0.375, 2.0) {};
		\node [style=none] (61)  at (-0.125, 2.0) {};
		\node [style=none] (62)  at (0.375, 2.5) {};
		\node [style=none] (63)  at (-0.125, 2.5) {};
		\node [style=none] (64)  at (0.0, -1.25) {};
		\node [style=none] (65)  at (0.75, -1.25) {};
		\node [style=none] (66)  at (0.0, -0.75) {};
		\node [style=none] (67)  at (0.75, -0.75) {};
		\node [style=none] (68)  at (0.375, -1.0) {$W_b$};
		\node [style=none] (69)  at (0.625, -1.25) {};
		\node [style=none] (70)  at (0.125, -1.25) {};
		\node [style=none] (71)  at (0.625, -0.75) {};
		\node [style=none] (72)  at (0.125, -0.75) {};
		\node [style=none] (73)  at (-1.625, -2.25) {};
		\node [style=none] (74)  at (1.125, -2.25) {};
		\node [style=none] (75)  at (-0.75, 2.5) {};
		\node [style=none] (76)  at (-0.75, -1.25) {};
		\node [style=none] (77)  at (-0.125, -1.25) {};
		\node [style=none] (78)  at (0.375, 3.0) {};
		\node [style=none] (79)  at (0.125, 1.75) {};
		\node [style=none] (80)  at (0.125, 1.25) {};
		\node [style=none] (81)  at (0.625, 1.25) {};
		\node [style=none] (82)  at (0.625, 1.75) {};
		\node [style=none] (83)  at (0.375, 1.25) {};
		\node [style=none] (84)  at (0.375, 1.75) {};
		\node [style=none] (85)  at (0.375, 1.5) {$B$};
		\node [style=none] (86)  at (1.75, 0.5) {};
		\node [style=none] (87)  at (2.5, 0.5) {};
		\node [style=none] (88)  at (1.75, 1.0) {};
		\node [style=none] (89)  at (2.5, 1.0) {};
		\node [style=none] (90)  at (2.125, 0.75) {$V_c$};
		\node [style=none] (91)  at (2.375, 0.5) {};
		\node [style=none] (92)  at (1.875, 0.5) {};
		\node [style=none] (93)  at (2.125, 1.0) {};
		\node [style=none] (94)  at (1.5, 2.0) {};
		\node [style=none] (95)  at (2.25, 2.0) {};
		\node [style=none] (96)  at (1.5, 2.5) {};
		\node [style=none] (97)  at (2.25, 2.5) {};
		\node [style=none] (98)  at (1.875, 2.25) {CX$_c$};
		\node [style=none] (99)  at (2.125, 2.0) {};
		\node [style=none] (100)  at (1.625, 2.0) {};
		\node [style=none] (101)  at (2.125, 2.5) {};
		\node [style=none] (102)  at (1.625, 2.5) {};
		\node [style=none] (103)  at (1.75, -1.25) {};
		\node [style=none] (104)  at (2.5, -1.25) {};
		\node [style=none] (105)  at (1.75, -0.75) {};
		\node [style=none] (106)  at (2.5, -0.75) {};
		\node [style=none] (107)  at (2.125, -1.0) {$W_c$};
		\node [style=none] (108)  at (2.375, -1.25) {};
		\node [style=none] (109)  at (1.875, -1.25) {};
		\node [style=none] (110)  at (2.375, -0.75) {};
		\node [style=none] (111)  at (1.875, -0.75) {};
		\node [style=none] (112)  at (-1.0, -2.25) {};
		\node [style=none] (113)  at (2.0, -2.25) {};
		\node [style=none] (114)  at (1.0, 2.5) {};
		\node [style=none] (115)  at (1.0, -1.25) {};
		\node [style=none] (116)  at (1.625, -1.25) {};
		\node [style=none] (117)  at (2.125, 3.0) {};
		\node [style=none] (118)  at (1.875, 1.75) {};
		\node [style=none] (119)  at (1.875, 1.25) {};
		\node [style=none] (120)  at (2.375, 1.25) {};
		\node [style=none] (121)  at (2.375, 1.75) {};
		\node [style=none] (122)  at (2.125, 1.25) {};
		\node [style=none] (123)  at (2.125, 1.75) {};
		\node [style=none] (124)  at (2.125, 1.5) {$C$};
		\node [style=none, font={\scriptsize}, text=black] (125)  at (-1.375, 3.25) {$\one\oplus\one$};
		\node [style=none, font={\scriptsize}, text=black] (126)  at (0.375, 3.25) {$\one\oplus\one$};
		\node [style=none, font={\scriptsize}, text=black] (127)  at (2.125, 3.25) {$\one\oplus\one$};
		\node [style=none] (131)  at (2.25, -2.75) {};
		\node [style=none] (132)  at (2.25, -3.0) {};
		\node [style=none, font={\scriptsize}, text=black] (133)  at (2.25, -3.25) {$\one\oplus\one$};
		\node [style=none] (134)  at (1.0, -2.75) {};
		\node [style=none] (135)  at (1.0, -3.0) {};
		\node [style=none, font={\scriptsize}, text=black] (136)  at (1.0, -3.25) {$\one\oplus\one$};
		\node [style=none] (218)  at (-1.325, -1.25) {};
		\node [style=none] (219)  at (0.05, -2.25) {};
		\node [style=none] (220)  at (0.425, -1.25) {};
		\node [style=none] (221)  at (0.925, -2.25) {};
		\node [style=none] (222)  at (2.175, -1.25) {};
		\node [style=none] (223)  at (1.8, -2.25) {};
		\node [style=none] (230)  at (-2.25, -2.75) {};
		\node [style=none] (231)  at (-2.25, -3.0) {};
		\node [style=none, font={\scriptsize}, text=black] (232)  at (-2.25, -3.25) {$\one\oplus(\one\oplus\one)$};
	\end{pgfonlayer}
	\begin{pgfonlayer}{edgelayer}
		\draw (0.center) to (2.center);
		\draw (1.center) to (0.center);
		\draw (2.center) to (3.center);
		\draw (3.center) to (1.center);
		\draw [draw=black, in=90, out=-90] (5.center) to (72.center);
		\draw [draw=black, in=90, out=-90, looseness=0.50] (6.center) to (110.center);
		\draw [draw=black] (7.center) to (43.center);
		\draw (8.center) to (10.center);
		\draw (9.center) to (8.center);
		\draw (10.center) to (11.center);
		\draw (11.center) to (9.center);
		\draw [draw=black] (13.center) to (44.center);
		\draw (17.center) to (19.center);
		\draw (18.center) to (17.center);
		\draw (19.center) to (20.center);
		\draw (20.center) to (18.center);
		\draw [draw=black, in=-270, out=-90] (22.center) to (34.center);
		\draw (26.center) to (27.center);
		\draw (27.center) to (29.center);
		\draw (28.center) to (26.center);
		\draw (29.center) to (28.center);
		\draw [black] (32.center) to (31.center);
		\draw [draw=black, in=-90, out=90] (33.center) to (37.center);
		\draw [draw=black, bend left=90, looseness=1.50] (35.center) to (16.center);
		\draw [draw=black, bend right=90, looseness=1.50] (36.center) to (23.center);
		\draw [draw=black] (36.center) to (35.center);
		\draw [draw=black] (37.center) to (14.center);
		\draw [draw=black] (38.center) to (15.center);
		\draw (39.center) to (42.center);
		\draw (40.center) to (39.center);
		\draw (40.center) to (41.center);
		\draw (42.center) to (41.center);
		\draw (47.center) to (49.center);
		\draw (48.center) to (47.center);
		\draw (49.center) to (50.center);
		\draw (50.center) to (48.center);
		\draw [draw=black, in=90, out=-90] (52.center) to (111.center);
		\draw [draw=black, in=90, out=-90] (53.center) to (24.center);
		\draw [draw=black] (54.center) to (83.center);
		\draw (55.center) to (57.center);
		\draw (56.center) to (55.center);
		\draw (57.center) to (58.center);
		\draw (58.center) to (56.center);
		\draw [draw=black] (60.center) to (84.center);
		\draw (64.center) to (66.center);
		\draw (65.center) to (64.center);
		\draw (66.center) to (67.center);
		\draw (67.center) to (65.center);
		\draw [draw=black, in=-270, out=-90] (69.center) to (74.center);
		\draw [draw=black, in=-90, out=90] (71.center) to (92.center);
		\draw [draw=black, in=-90, out=90] (73.center) to (77.center);
		\draw [draw=black, bend left=90, looseness=1.50] (75.center) to (63.center);
		\draw [draw=black, bend right=90, looseness=1.50] (76.center) to (70.center);
		\draw [draw=black] (76.center) to (75.center);
		\draw [draw=black] (77.center) to (61.center);
		\draw [draw=black] (78.center) to (62.center);
		\draw (79.center) to (82.center);
		\draw (80.center) to (79.center);
		\draw (80.center) to (81.center);
		\draw (82.center) to (81.center);
		\draw (86.center) to (88.center);
		\draw (87.center) to (86.center);
		\draw (88.center) to (89.center);
		\draw (89.center) to (87.center);
		\draw [draw=black, in=90, out=-90, looseness=0.50] (91.center) to (25.center);
		\draw [draw=black] (93.center) to (122.center);
		\draw (94.center) to (96.center);
		\draw (95.center) to (94.center);
		\draw (96.center) to (97.center);
		\draw (97.center) to (95.center);
		\draw [draw=black] (99.center) to (123.center);
		\draw (103.center) to (105.center);
		\draw (104.center) to (103.center);
		\draw (105.center) to (106.center);
		\draw (106.center) to (104.center);
		\draw [draw=black, in=90, out=-90, looseness=0.75] (108.center) to (113.center);
		\draw [draw=black, in=-90, out=90, looseness=0.50] (112.center) to (116.center);
		\draw [draw=black, bend left=90, looseness=1.50] (114.center) to (102.center);
		\draw [draw=black, bend right=90, looseness=1.50] (115.center) to (109.center);
		\draw [draw=black] (115.center) to (114.center);
		\draw [draw=black] (116.center) to (100.center);
		\draw [draw=black] (117.center) to (101.center);
		\draw (118.center) to (121.center);
		\draw (119.center) to (118.center);
		\draw (119.center) to (120.center);
		\draw (121.center) to (120.center);
		\draw [black] (132.center) to (131.center);
		\draw [black] (135.center) to (134.center);
		\draw [draw=black, in=-270, out=-90] (218.center) to (219.center);
		\draw [draw=black, in=-270, out=-90] (220.center) to (221.center);
		\draw [draw=black, in=90, out=-90, looseness=0.75] (222.center) to (223.center);
		\draw [black] (231.center) to (230.center);
	\end{pgfonlayer}
\end{tikzpicture}

%% file: figures/Lugano-2/Lugano-2_01.tikz
\begin{tikzpicture}
	\begin{pgfonlayer}{nodelayer}
		\node [style=none] (138)  at (-2.875, -2.825) {};
		\node [style=none] (139)  at (-2.875, -2.325) {};
		\node [style=none] (140)  at (2.875, -2.825) {};
		\node [style=none] (141)  at (2.875, -2.325) {};
		\node [style=none] (142)  at (-0.125, -2.575) {$F$};
		\node [style=none] (145)  at (-1.5, 1.675) {};
		\node [style=none] (146)  at (-1.5, 1.175) {};
		\node [style=none] (147)  at (-1.0, 1.175) {};
		\node [style=none] (148)  at (-1.0, 1.675) {};
		\node [style=none] (149)  at (-1.25, 1.425) {$A$};
		\node [style=none] (151)  at (0.25, 1.675) {};
		\node [style=none] (152)  at (0.25, 1.175) {};
		\node [style=none] (153)  at (0.75, 1.175) {};
		\node [style=none] (154)  at (0.75, 1.675) {};
		\node [style=none] (155)  at (0.5, 1.425) {$B$};
		\node [style=none] (156)  at (2.0, 1.675) {};
		\node [style=none] (157)  at (2.0, 1.175) {};
		\node [style=none] (158)  at (2.5, 1.175) {};
		\node [style=none] (159)  at (2.5, 1.675) {};
		\node [style=none] (160)  at (2.25, 1.425) {$C$};
		\node [style=contraction] (167)  at (-1.25, 2.025) {};
		\node [style=none] (168)  at (-1.25, 1.675) {};
		\node [style=contraction] (169)  at (-1.25, 2.975) {};
		\node [style=none] (170)  at (-1.25, 3.325) {};
		\node [style=box] (171)  at (-0.975, 2.5) {$X$};
		\node [style=plus] (172)  at (-1.25, 0.825) {};
		\node [style=none] (173)  at (-1.25, 1.175) {};
		\node [style=dot] (174)  at (-1.625, 0.325) {};
		\node [style=dot] (175)  at (-0.875, 0.325) {};
		\node [style=plus] (176)  at (0.5, 0.825) {};
		\node [style=none] (177)  at (0.5, 1.175) {};
		\node [style=dot] (178)  at (0.125, 0.325) {};
		\node [style=dot] (179)  at (0.875, 0.325) {};
		\node [style=plus] (180)  at (2.25, 0.825) {};
		\node [style=none] (181)  at (2.25, 1.175) {};
		\node [style=dot] (182)  at (1.875, 0.325) {};
		\node [style=dot] (183)  at (2.625, 0.325) {};
		\node [style=plus] (184)  at (0.125, -1.45) {};
		\node [style=plus] (185)  at (0.625, -1.45) {};
		\node [style=plus] (186)  at (-2.625, -1.975) {};
		\node [style=tensor] (187)  at (-2.875, -1.5) {};
		\node [style=tensor] (188)  at (-2.375, -1.5) {};
		\node [style=none] (189)  at (-2.625, -2.325) {};
		\node [style=plus] (192)  at (1.125, -1.45) {};
		\node [style=plus] (193)  at (1.625, -1.45) {};
		\node [style=plus] (194)  at (-1.625, -1.975) {};
		\node [style=tensor] (195)  at (-1.875, -1.5) {};
		\node [style=tensor] (196)  at (-1.375, -1.5) {};
		\node [style=none] (197)  at (-1.625, -2.325) {};
		\node [style=plus] (200)  at (2.125, -1.45) {};
		\node [style=plus] (201)  at (2.625, -1.45) {};
		\node [style=plus] (202)  at (-0.625, -1.975) {};
		\node [style=tensor] (203)  at (-0.875, -1.5) {};
		\node [style=tensor] (204)  at (-0.375, -1.5) {};
		\node [style=none] (205)  at (-0.625, -2.325) {};
		\node [style=contraction] (208)  at (0.5, 2.025) {};
		\node [style=none] (209)  at (0.5, 1.675) {};
		\node [style=contraction] (210)  at (0.5, 2.975) {};
		\node [style=none] (211)  at (0.5, 3.325) {};
		\node [style=box] (212)  at (0.775, 2.5) {$X$};
		\node [style=contraction] (213)  at (2.25, 2.025) {};
		\node [style=none] (214)  at (2.25, 1.675) {};
		\node [style=contraction] (215)  at (2.25, 2.975) {};
		\node [style=none] (216)  at (2.25, 3.325) {};
		\node [style=box] (217)  at (2.525, 2.5) {$X$};
		\node [style=none] (224)  at (0.125, -2.325) {};
		\node [style=none] (225)  at (0.625, -2.325) {};
		\node [style=none] (226)  at (1.125, -2.325) {};
		\node [style=none] (227)  at (1.625, -2.325) {};
		\node [style=none] (228)  at (2.125, -2.325) {};
		\node [style=none] (229)  at (2.625, -2.325) {};
		\node [style=none] (233)  at (-0.375, -2.825) {};
		\node [style=none] (234)  at (-0.375, -3.075) {};
		\node [style=none, font={\scriptsize}, text=black] (235)  at (-0.375, -3.325) {$\one\oplus\one$};
		\node [style=none] (236)  at (2.625, -2.825) {};
		\node [style=none] (237)  at (2.625, -3.075) {};
		\node [style=none, font={\scriptsize}, text=black] (238)  at (2.625, -3.325) {$\one\oplus\one$};
		\node [style=none] (239)  at (1.125, -2.825) {};
		\node [style=none] (240)  at (1.125, -3.075) {};
		\node [style=none, font={\scriptsize}, text=black] (241)  at (1.125, -3.325) {$\one\oplus\one$};
		\node [style=none] (242)  at (-2.625, -2.825) {};
		\node [style=none] (243)  at (-2.625, -3.075) {};
		\node [style=none, font={\scriptsize}, text=black] (244)  at (-2.625, -3.325) {$\one\oplus(\one\oplus\one)$};
	\end{pgfonlayer}
	\begin{pgfonlayer}{edgelayer}
		\draw (138.center) to (139.center);
		\draw (139.center) to (141.center);
		\draw (140.center) to (138.center);
		\draw (141.center) to (140.center);
		\draw (145.center) to (148.center);
		\draw (146.center) to (145.center);
		\draw (146.center) to (147.center);
		\draw (148.center) to (147.center);
		\draw (151.center) to (154.center);
		\draw (152.center) to (151.center);
		\draw (152.center) to (153.center);
		\draw (154.center) to (153.center);
		\draw (156.center) to (159.center);
		\draw (157.center) to (156.center);
		\draw (157.center) to (158.center);
		\draw (159.center) to (158.center);
		\draw [style=compact dash, violet, bend left=45, looseness=1.25] (167) to (169);
		\draw [style=compact dash, blue, bend right=45, looseness=1.25] (167) to (169);
		\draw (168.center) to (167);
		\draw (170.center) to (169);
		\draw [violet] (172) to (174);
		\draw [blue] (172) to (175);
		\draw (173.center) to (172);
		\draw [violet, in=150, out=-135, looseness=0.75] (174) to (192);
		\draw [blue, in=60, out=-90, looseness=0.75] (175) to (192);
		\draw [blue, in=150, out=-30] (175) to (201);
		\draw [cyan] (176) to (178);
		\draw [green] (176) to (179);
		\draw (177.center) to (176);
		\draw [cyan, in=45, out=-135] (178) to (185);
		\draw [cyan, in=150, out=-30] (178) to (200);
		\draw [orange] (180) to (182);
		\draw [red] (180) to (183);
		\draw (181.center) to (180);
		\draw [orange, in=-135, out=135, looseness=0.75] (184) to (182);
		\draw [red] (184) to (183);
		\draw [green, in=225, out=135] (185) to (179);
		\draw [style=compact dash, violet, in=150, out=-90] (187) to (186);
		\draw [blue, style=compact dash, in=30, out=-90] (188) to (186);
		\draw (189.center) to (186);
		\draw [orange, in=-60, out=45] (193) to (182);
		\draw [red, in=-75, out=120] (193) to (183);
		\draw [style=compact dash, cyan, in=150, out=-90] (195) to (194);
		\draw [green, style=compact dash, in=30, out=-90] (196) to (194);
		\draw (197.center) to (194);
		\draw [green, in=-45, out=45, looseness=0.75] (200) to (179);
		\draw [violet, in=-30, out=45, looseness=0.75] (201) to (174);
		\draw [style=compact dash, orange, in=150, out=-90] (203) to (202);
		\draw [red, style=compact dash, in=30, out=-90] (204) to (202);
		\draw (205.center) to (202);
		\draw [style=compact dash, cyan, bend left=45, looseness=1.25] (208) to (210);
		\draw [style=compact dash, green, bend right=45, looseness=1.25] (208) to (210);
		\draw (209.center) to (208);
		\draw (211.center) to (210);
		\draw [style=compact dash, orange, bend left=45, looseness=1.25] (213) to (215);
		\draw [style=compact dash, red, bend right=45, looseness=1.25] (213) to (215);
		\draw (214.center) to (213);
		\draw (216.center) to (215);
		\draw (224.center) to (184);
		\draw (225.center) to (185);
		\draw (226.center) to (192);
		\draw (227.center) to (193);
		\draw (228.center) to (200);
		\draw (229.center) to (201);
		\draw [black] (234.center) to (233.center);
		\draw [black] (237.center) to (236.center);
		\draw [black] (240.center) to (239.center);
		\draw [black] (243.center) to (242.center);
	\end{pgfonlayer}
\end{tikzpicture}

%% file: figures/Lugano-2/Lugano-2_02.tikz
\begin{tikzpicture}
	\begin{pgfonlayer}{nodelayer}
		\node [style=none] (251)  at (-0.7, 0.35) {};
		\node [style=none] (252)  at (-0.7, -0.15) {};
		\node [style=none] (253)  at (-0.2, -0.15) {};
		\node [style=none] (254)  at (-0.2, 0.35) {};
		\node [style=none] (255)  at (-0.45, 0.1) {$A$};
		\node [style=none] (256)  at (0.3, 0.35) {};
		\node [style=none] (257)  at (0.3, -0.15) {};
		\node [style=none] (258)  at (0.8, -0.15) {};
		\node [style=none] (259)  at (0.8, 0.35) {};
		\node [style=none] (260)  at (0.55, 0.1) {$B$};
		\node [style=none] (261)  at (1.3, 0.35) {};
		\node [style=none] (262)  at (1.3, -0.15) {};
		\node [style=none] (263)  at (1.8, -0.15) {};
		\node [style=none] (264)  at (1.8, 0.35) {};
		\node [style=none] (265)  at (1.55, 0.1) {$C$};
		\node [style=contraction] (266)  at (-0.45, 0.7) {};
		\node [style=none] (267)  at (-0.45, 0.35) {};
		\node [style=contraction] (268)  at (-0.45, 1.65) {};
		\node [style=none] (269)  at (-0.45, 2.0) {};
		\node [style=box] (270)  at (-0.175, 1.175) {$X$};
		\node [style=none] (272)  at (-0.45, -0.15) {};
		\node [style=none] (276)  at (0.55, -0.15) {};
		\node [style=none] (280)  at (1.55, -0.15) {};
		\node [style=plus] (285)  at (-1.575, -1.4) {};
		\node [style=tensor] (287)  at (-1.825, -0.925) {};
		\node [style=none] (288)  at (-1.575, -1.75) {};
		\node [style=tensor] (293)  at (-1.575, -0.425) {};
		\node [style=plus] (297)  at (-1.325, -0.9) {};
		\node [style=tensor] (299)  at (-1.075, -0.425) {};
		\node [style=contraction] (301)  at (0.55, 0.7) {};
		\node [style=none] (302)  at (0.55, 0.35) {};
		\node [style=contraction] (303)  at (0.55, 1.65) {};
		\node [style=none] (304)  at (0.55, 2.0) {};
		\node [style=box] (305)  at (0.825, 1.175) {$X$};
		\node [style=contraction] (306)  at (1.55, 0.7) {};
		\node [style=none] (307)  at (1.55, 0.35) {};
		\node [style=contraction] (308)  at (1.55, 1.65) {};
		\node [style=none] (309)  at (1.55, 2.0) {};
		\node [style=box] (310)  at (1.825, 1.175) {$X$};
		\node [style=none, font={\scriptsize}, text=black] (328)  at (-1.575, -2.0) {$\one\oplus(\one\oplus\one)$};
		\node [style=none, font={\scriptsize}, text=black] (329)  at (-0.45, -2.0) {$\one\oplus\one$};
		\node [style=none, font={\scriptsize}, text=black] (330)  at (0.55, -2.0) {$\one\oplus\one$};
		\node [style=none, font={\scriptsize}, text=black] (331)  at (1.55, -2.0) {$\one\oplus\one$};
		\node [style=plus] (332)  at (-0.45, -1.35) {};
		\node [style=none] (333)  at (-0.45, -1.75) {};
		\node [style=plus] (334)  at (-0.45, -0.5) {};
		\node [style=plus] (335)  at (0.55, -0.5) {};
		\node [style=plus] (336)  at (0.55, -1.35) {};
		\node [style=none] (337)  at (0.55, -1.75) {};
		\node [style=plus] (338)  at (1.55, -0.5) {};
		\node [style=plus] (339)  at (1.55, -1.35) {};
		\node [style=none] (340)  at (1.55, -1.75) {};
	\end{pgfonlayer}
	\begin{pgfonlayer}{edgelayer}
		\draw (251.center) to (254.center);
		\draw (252.center) to (251.center);
		\draw (252.center) to (253.center);
		\draw (254.center) to (253.center);
		\draw (256.center) to (259.center);
		\draw (257.center) to (256.center);
		\draw (257.center) to (258.center);
		\draw (259.center) to (258.center);
		\draw (261.center) to (264.center);
		\draw (262.center) to (261.center);
		\draw (262.center) to (263.center);
		\draw (264.center) to (263.center);
		\draw [style=compact dash, violet, bend left=45, looseness=1.25] (266) to (268);
		\draw [style=compact dash, blue, bend right=45, looseness=1.25] (266) to (268);
		\draw (267.center) to (266);
		\draw (269.center) to (268);
		\draw [blue, style=compact dash, in=150, out=-90] (287) to (285);
		\draw (288.center) to (285);
		\draw [green, style=compact dash, in=150, out=-90] (293) to (297);
		\draw [style=compact dash, violet, in=30, out=-90] (297) to (285);
		\draw [red, style=compact dash, in=30, out=-90] (299) to (297);
		\draw [style=compact dash, cyan, bend left=45, looseness=1.25] (301) to (303);
		\draw [style=compact dash, green, bend right=45, looseness=1.25] (301) to (303);
		\draw (302.center) to (301);
		\draw (304.center) to (303);
		\draw [style=compact dash, orange, bend left=45, looseness=1.25] (306) to (308);
		\draw [style=compact dash, red, bend right=45, looseness=1.25] (306) to (308);
		\draw (307.center) to (306);
		\draw (309.center) to (308);
		\draw [violet, bend left=45, looseness=1.25] (332) to (334);
		\draw [blue, bend right=45, looseness=1.25] (332) to (334);
		\draw (333.center) to (332);
		\draw (334) to (272.center);
		\draw [cyan, bend right=45, looseness=1.25] (335) to (336);
		\draw [green, bend left=45, looseness=1.25] (335) to (336);
		\draw (335) to (276.center);
		\draw (337.center) to (336);
		\draw [orange, bend right=45, looseness=1.25] (338) to (339);
		\draw [red, bend left=45, looseness=1.25] (338) to (339);
		\draw (338) to (280.center);
		\draw (340.center) to (339);
	\end{pgfonlayer}
\end{tikzpicture}